\documentclass[aps,prd,reprint,twocolumn,superscriptaddress,showpacs]{revtex4-1}

\usepackage{graphicx}
\usepackage{mathrsfs}
\usepackage{bm}
\usepackage{amsmath}
\usepackage{dcolumn}
\usepackage{epstopdf}
\usepackage{dsfont}
\usepackage{amssymb}
\usepackage{tabularx}
\usepackage{array}
\usepackage{float}
\usepackage{color}
\usepackage{epstopdf}
\usepackage{mathrsfs}
\usepackage[colorlinks, linkcolor=blue,anchorcolor=blue,citecolor=blue,urlcolor=blue]{hyperref}

\begin{document}
\title{Tunable anomalous Hall transport in bulk and two-dimensional 1\emph{T}-CrTe$_{2}$: A first-principles study}
	
\author{Si Li}
\affiliation{School of Physics, Northwest University, Xi'an 710069, China}
\affiliation{Research Laboratory for Quantum Materials, Singapore University of Technology and Design, Singapore 487372, Singapore}
\affiliation{School of Physics and Electronics, Hunan Normal University, Changsha, Hunan 410081, China}

\author{Shan-Shan Wang}
\affiliation{School of Physics, Southeast University, Nanjing 211189, China}
\affiliation{Research Laboratory for Quantum Materials, Singapore University of Technology and Design, Singapore 487372, Singapore}

\author{Bo Tai}
\affiliation{Research Laboratory for Quantum Materials, Singapore University of Technology and Design, Singapore 487372, Singapore}

\author{Weikang Wu}
\affiliation{Research Laboratory for Quantum Materials, Singapore University of Technology and Design, Singapore 487372, Singapore}

\author{Bin Xiang}
\affiliation{Hefei National Research Center for Physical Sciences at the Microscale, Department of Materials Science \& Engineering,
CAS Key Lab of Materials for Energy Conversion, University of Science and Technology of China, Hefei 230026, China}

\author{Xian-Lei Sheng}
\affiliation{School of Physics, and Key Laboratory of Micro-nano Measurement-Manipulation and Physics, Beihang University, Beijing 100191, China}

\author{Shengyuan A. Yang}
%\email{shengyuan\_yang@sutd.edu.sg}
\affiliation{Research Laboratory for Quantum Materials, Singapore University of Technology and Design, Singapore 487372, Singapore}

\begin{abstract}
Layered materials with robust magnetic ordering have been attracting significant research interest. In recent experiments, a new layered material 1$T$-CrTe$_{2}$ has been synthesized and exhibits ferromagnetism above the room temperature.
Here, based on first-principles calculations, we investigate the electronic, magnetic, and transport properties of 1$T$-CrTe$_{2}$, both in the bulk and in the two-dimensional (2D) limit. We show that 1$T$-CrTe$_{2}$ can be stable in the monolayer form, and has a low exfoliation energy. The monolayer structure is an intrinsic ferromagnetic metal, which maintains a relatively high Curie temperature above 200 K.
Particularly, we reveal interesting features in the anomalous Hall transport. We show that in the ground state, both bulk and monolayer 1$T$-CrTe$_{2}$ possess vanishing anomalous Hall effect, because the magnetization preserves one vertical mirror symmetry. The anomalous Hall conductivity can be made sizable by tuning the magnetization direction or by uniaxial strains that break the mirror symmetry.
\end{abstract}
	
\maketitle
\section{Introduction}

Layered materials have been attracting great interest in recent research, because they offer a feasible route to achieve high-quality two-dimensional (2D) materials~\cite{das2015beyond,bhimanapati2015recent,choi2017recent}. This has been well demonstrated in the cases of graphene~\cite{novoselov2004electric}, 2D MoS$_2$~\cite{novoselov2005two,mak2010atomically}, and phosphorene~\cite{liu2014phosphorene,li2014black}, which were all first obtained from their layered bulk via exfoliation method.

One focus in the field is to introduce magnetic ordering into 2D materials, which is intriguing from both fundamental and application perspectives. In 2017, the first two 2D intrinsic magnetic materials were reported in 2D CrI$_3$ and Cr$_2$Ge$_2$Te$_6$, which are ferromagnetic (FM) with Curie temperatures of 45 K and 66 K, respectively~\cite{huang2017layer,gong2017discovery}. Subsequently, several other 2D magnetic systems were also identified, such as Fe$_3$GeTe$_2$~\cite{zhu2016electronic,zhuang2016strong,deng2018gate}, VSe$_2$~\cite{bonilla2018strong}, VTe$_2$~\cite{sugawara2019monolayer}, and MnSe$_x$~\cite{o2018room}.
It was demonstrated that devices made from such 2D magnetic materials could achieve superior performances~\cite{wang2018electric,deng2018gate,sun2019giant}, such as ultrahigh magnetoresistance~\cite{song2018giant} and efficient current-induced magnetic switching~\cite{wang2019current}, which are highly promising for nanoscale device applications. Currently, the family of 2D magnetic materials is still quite limited. It is hence an urgent task to explore new candidates, especially those with high transition temperatures and unique properties derived from magnetism.

To achieve 2D magnetic materials, a natural approach is to look for layered materials with robust magnetic ordering in the bulk, then check whether the magnetism can be maintained in the 2D limit. In fact, 2D CrI$_3$ and Cr$_2$Ge$_2$Te$_6$ were both discovered in this approach.
Recently, a new layered compound, 1$T$-CrTe$_{2}$, was successfully synthesized in experiment, and it was found to be a FM metal with a high Curie temperature of $310$ $\mathrm{K}$~\cite{freitas2015ferromagnetism}. The discovery attracted immediate interest. For example, Lv \emph{et al.}~\cite{lv2015strain} theoretically studied the strain effects on the magnetism in monolayer CrTe$_{2}$ and revealed an interesting magnetic phase transition at compressive strains. Sui \emph{et al.}~\cite{sui2017voltage} calculated the electric field effect on the magnetic anisotropy for a large family of 2D transition metal dichalcogenides, including monolayer CrTe$_{2}$. Most recently, the experimental study by Sun \emph{et al.}~\cite{sun2019room} reported that the room-temperature FM ordering can be maintained in few-layer (down to 5 layers) 1$T$-CrTe$_{2}$.

Motivated by these recent progress,
in this work, based on first-principles calculations, we systematically investigate the electronic, magnetic, and transport properties of 1$T$-CrTe$_{2}$, in both bulk and 2D monolayer limits. We show that the monolayer 1$T$-CrTe$_{2}$ is dynamically stable, and has a low exfoliation energy, comparable to other existing 2D materials. The FM ordering is robust in the monolayer limit. Our Monte Carlo simulation shows that the Curie temperature for monolayer is still quite high $> 200$ K. For both bulk and monolayer, we find that the magnetization prefers an in-plane direction that preserves a vertical mirror. As a result, the anomalous Hall effect is fully suppressed in the ground state of these systems, and can only be made non-vanishing by breaking this mirror symmetry. We discuss two methods to turn on the anomalous Hall transport, by re-orienting the magnetization (e.g., by an applied magnetic field) and by applying an uniaxial strain. These are verified by our calculation of the intrinsic anomalous Hall conductivity, of which the obtained values are comparable to typical transition metal ferromagnets. Our results provide useful guidance to explore an intriguing magnetic material. The robust ferromagnetism and the sensitive dependence of transport (on external field and strain) will open opportunities for the design of novel sensors and functional devices at nanoscale.

\section{Computation Method}

Our first-principles calculations were based on the density functional theory (DFT), using the projector augmented wave method as implemented in the Vienna \emph{ab initio} simulation package~\cite{Kresse1994,Kresse1996,PAW}. The generalized gradient
approximation with the Perdew-Burke-Ernzerhof (PBE)~\cite{PBE} realization was adopted for the exchange-correlation functional. The cutoff energy was set as 450 eV. The energy and force convergence criteria were set to be $10^{-7}$ eV and $0.01$ eV/\AA, respectively. For the calculation of the bulk, we adopted the experimental lattice parameters and a $16\times 16\times 8$ $\Gamma$-centered $k$-point mesh. The van der Waals (vdW) corrections have been taken into
account by the approach of Dion \emph{et al.}~\cite{Dion2004}. For the calculation of the monolayer, a $16\times 16\times 1$ $\Gamma$-centered $k$-point mesh was used and a vacuum layer with a thickness of 20 \AA\, was taken to avoid artificial interactions between periodic images. The phonon spectrum was calculated using the PHONOPY code through the DFPT approach~\cite{Togo2015}. To account for the correlation effects for the Cr-3$d$ orbitals, the DFT$+U$ method~\cite{Anisimov1991,dudarev1998} was used for calculating the band structures. For the results presented in the main text, the $U$ value was taken to be 2 eV~\cite{sui2017voltage,lado2017origin}. The test results of other $U$ values are presented in the Supplemental Material~\cite{SM}. The Berry curvature and intrinsic anomalous Hall conductivity were evaluated on a denser $k$-mesh of $400 \times 400 \times 1$ for the monolayer and $100 \times 100 \times 100$ for the bulk material using the WANNIER90 package~\cite{mostofi2008wannier90,wang2006ab}.

\section{CRYSTAL STRUCTURE}
\begin{figure}[b]
	\includegraphics[width=8.6cm]{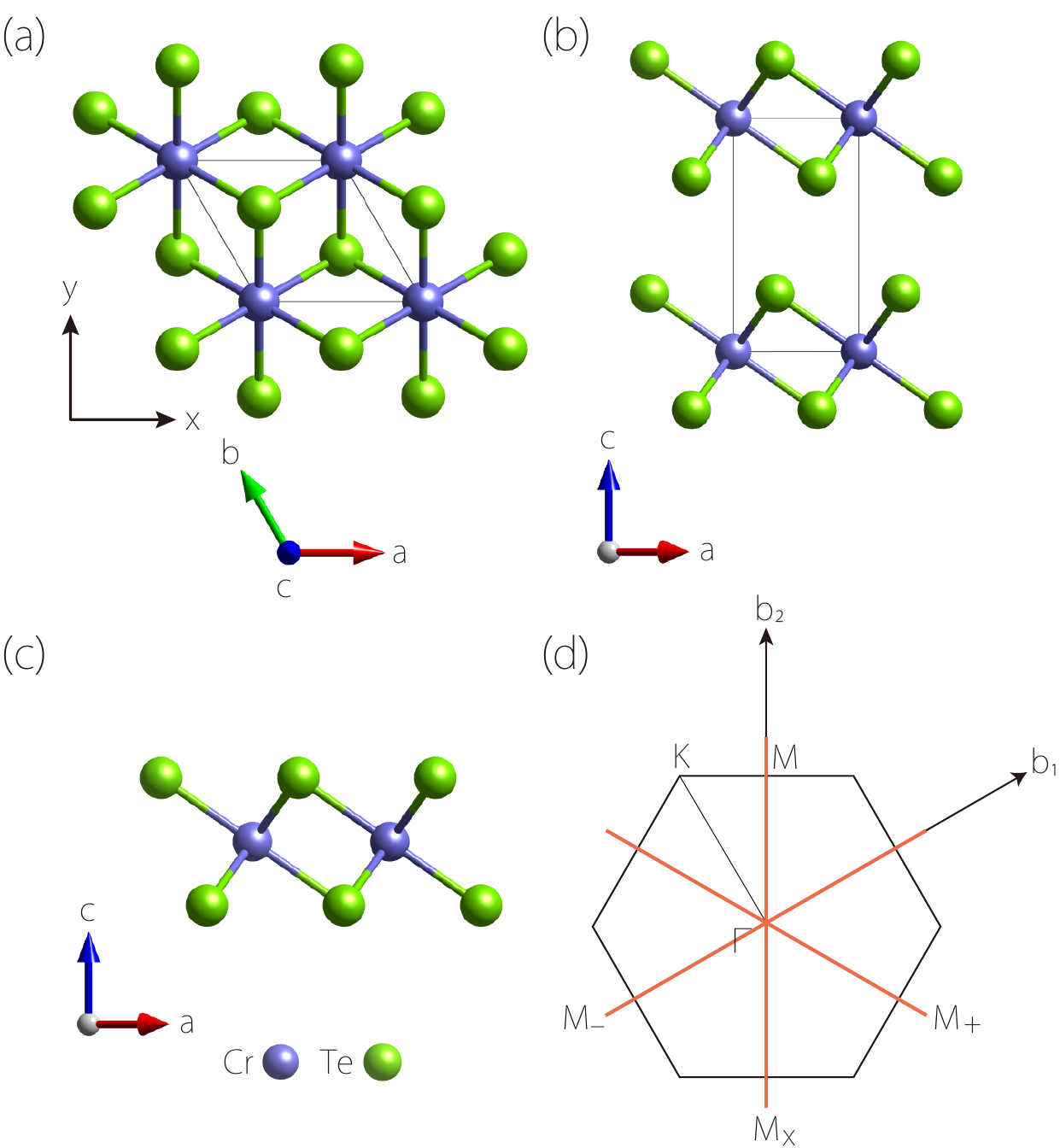}
	\caption{ (a) Top view and (b) side view of the crystal structure of bulk 1$T$-CrTe$_{2}$. The primitive cell is shown with the solid line. (c) Crystal structure of the monolayer 1$T$-CrTe$_{2}$. (d) Brillouin zone for monolayer 1$T$-CrTe$_{2}$ with high symmetry points labeled. The three vertical mirrors for the lattice structure are marked by the red lines.}
	\label{fig1}
\end{figure}

Single crystals of 1$T$-CrTe$_{2}$ have been synthesized by oxidation of KCrTe$_2$~\cite{freitas2015ferromagnetism}. The bulk material has a layered trigonal CdI$_2$-type crystal structure, with space group $P \overline{3} m 1$ (No.~$164$). As shown in Fig.~\ref{fig1}(a) and (b), each CrTe$_{2}$ layer consists of Cr atomic layer sandwiched by two Te atomic layers, such that each Cr atom is surrounded by six Te atoms, forming an octahedral crystal field. The structural data have been fully determined by the X-ray powder diffraction method. The experimental values of the lattice parameters are given by $a =b= 3.7887$ \AA, and $c = 6.0955$ \AA~\cite{freitas2015ferromagnetism}.

In Fig.~\ref{fig1}(c), we isolate one monolayer CrTe$_{2}$ from the bulk structure. It is found that the monolayer shares the same space group symmetry as the bulk. The point group symmetry for the structure (and also bulk) is $D_{3d}$, with generators of a rotoreflection $S_{6}$ and a vertical mirror $M_{x}$. Combining these two operations leads to another two mirrors and the inversion symmetry $\mathcal{P}$. The vertical mirror symmetry $M_{x}$ plays an important role in our later discussion of the anomalous Hall effect, so it is highlighted in Fig.~\ref{fig1}(d). From our first-principles calculation, the fully optimized lattice parameters for the monolayer are given by $a =b= 3.729$ \AA.

\section{Stability and Exfoliation Energy}

\begin{figure}[b]
	\includegraphics[width=8.6cm]{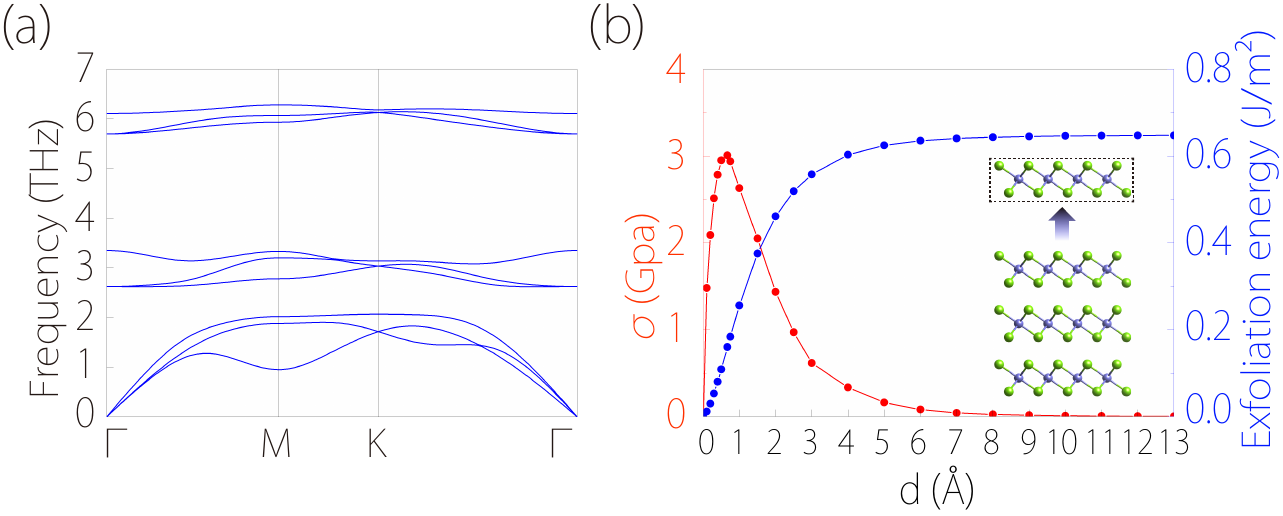}
	\caption{(a) Phonon spectrum for monolayer 1$T$-CrTe$_{2}$.(b) Exfoliation energy (blue line) for 1$T$-CrTe$_{2}$ as a function of its separation distance $d$ from the bulk (as illustrated in the inset). Here the bulk is modeled by three 1$T$-CrTe$_{2}$ layers in the calculation. The red curve shows the exfoliation strength $\sigma$ (i.e., the derivative of exfoliation energy with respect to $d$).}
	\label{fig2}
\end{figure}

To confirm the stability of the monolayer structure, we perform the phonon spectrum calculation. The obtained phonon spectrum is plotted in Fig.~\ref{fig2}(a),  which shows that there is no soft mode throughout the Brillouin zone (BZ), indicating that the structure is dynamically stable. The group velocities for the acoustic phonon branches are about $2.6 \times 10^{5}$ $\mathrm{cm} / \mathrm{s}$.

To assess the feasibility to obtain the monolayer CrTe$_{2}$ by exfoliation method, we evaluate its exfoliation energy. This is done by calculating the energy variation ($\delta E$) when a single monolayer is separated from the bulk by a distance $d$ (which simulates the exfoliation process, as illustrated in Fig.~\ref{fig2}(b)). With increasing $d$, the energy saturates to a value corresponding to the exfoliation energy. As shown in Fig.~\ref{fig2}(b), the exfoliation energy obtained from our calculation is about 0.647 J/m$^2$. This value is comparable to that of graphene (0.37 J/m$^2$)~\cite{zacharia2004} and MoS$_2$ (0.41 J/m$^2$), and is less than that of Ca$_2$N (1.14 J/m$^2$)~\cite{zhao2014,guan2015}. We have also calculated the exfoliation strength $\sigma$, which is defined as the maximum derivative of $\delta E$ with respect to the separation $d$. The obtained exfoliation strength is about 3.0 GPa, also similar to the values for typical 2D materials, such
as graphene ($\sim$2.1 GPa)~\cite{zhao2014}. These results indicate that monolayer 1$T$-CrTe$_{2}$ can be readily obtained from the bulk material by mechanical exfoliation. The recent experiment in Ref.~\cite{sun2019room} indeed achieved few-layer 1$T$-CrTe$_{2}$ by using the exfoliation method.

\section{Magnetic ordering}

\begin{figure}[b]
	\includegraphics[width=7.2cm]{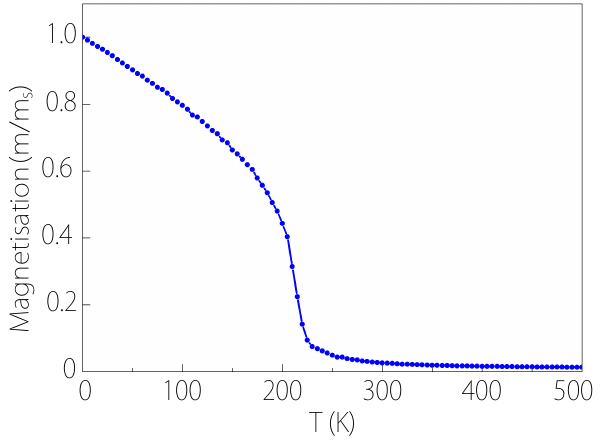}
	\caption{The normalized magnetic moment of monolayer 1$T$-CrTe$_{2}$ as a function of temperature by Monte Carlo simulations.}
	\label{fig3}
\end{figure}

The transition metal element Cr often brings about magnetism in its compounds. In 1$T$-CrTe$_{2}$, as we have mentioned, each Cr ion is sitting inside the octahedral crystal field formed by six neighboring Te ions. The crystal field splits the Cr 3$d$ orbitals into $e_g$ and $t_{2g}$ groups, with $e_g$ having a higher energy.  With the $d^2$ electron count for the Cr$^{4+}$ ion, according to the Hund's rule,
only the low-lying $t_{2g}$ orbitals are partially occupied
with parallel spins. This indicates that the Cr ions in 1$T$-CrTe$_{2}$ carry magnetic moments. In addition, for partially occupied $t_{2g}$ orbitals, the spin-orbit coupling (SOC) is typically much stronger compared to the $e_g$ case. {This is because the  orbital angular momentum is completely quenched for $e_g$ but not for $t_{2g}$. Hence, the SOC effects can appear at first order for partially filled $t_{2g}$ bands.} This could help to enhance the magnetic anisotropy and the Curie temperature.

The magnetic ordering of bulk 1$T$-CrTe$_{2}$ has been determined in previous experiment~\cite{freitas2015ferromagnetism}, which is FM and the magnetization lies in the layer plane. The Curie temperature is about 310 K. In the following, we focus on the magnetic ordering in monolayer CrTe$_{2}$.

We find that the monolayer CrTe$_{2}$ again prefers a FM ground state, which agrees with the calculation result in Ref.~\cite{lv2015strain} and the experimental result obtained for few-layer CrTe$_{2}$~\cite{sun2019room}. In addition, we find that the ground-state magnetization is in-plane and perpendicular to a vertical mirror plane (remember there are three equivalent mirrors). The magneto-anisotropy energy for the out-of-plane configuration is very high, reaching above $10^3$ $\mu$eV per Cr. This is much higher than that of typical 3$d$ transition metal ferromagnets, such as Fe (1.4 $\mu$eV per atom), Co (65 $\mu$eV per atom) and Ni (2.7 $\mu$eV per atom)~\cite{daalderop1988magnetic,lehnert2010magnetic}, {and is comparable to FePt and CoPt type alloys~\cite{sakuma1994first,ravindran2001large,shick2003coulomb}.} Meanwhile, the in-plane magnetic anisotropy is relatively small, $\sim 3$ $\mu$eV per Cr. Most importantly, the spontaneous magnetic ordering preserves one of the vertical mirrors, taken as $M_x$ here. This will have important consequences on the anomalous Hall effect.

We have estimated the Curie temperature ($T_C$) for the monolayer CrTe$_{2}$. The calculation is done by using the Monte Carlo simulation approach based on a classical effective spin model~\cite{evans2014atomistic}:
\begin{equation}\label{Heisenberg}
 H=-\sum_{ i, j} J_{i j} \bm{S}_{i} \cdot \bm{S}_{j}+K\sum_{i}\left(S_{i}^{z}\right)^{2},
\end{equation}
where $\bm S_i$ is the normalized spin vector on site $i$, $J_{ij}$ is the exchange coupling constant between sites $i$ and $j$, and $K$ is the anisotropy strength. {We include the nearest-neighbor coupling $J_1$ and the second-neighbor coupling $J_2$ in the model. These exchange coupling parameters may be obtained by several possible approaches, such as the energy mapping method, the strong-coupling perturbation expansion, and the local force approach~\cite{liechtenstein1987local,fischer2009exchange,nomura2020magnetic}. Here, we adopt the widely used energy mapping method (see the Supplemental Material for details~\cite{SM}). The obtained model parameters are given by $J_1=2.12 \times 10^{-21}$ J, $J_2=3.82 \times 10^{-22}$ J, and $K= 1.66 \times 10^{-22}$ J.
The Curie temperature is obtained from the variation of the net magnetization with respect to the temperature, which is shown in Fig.~\ref{fig3}. The estimated $T_C$ value is about 215 K. It should be noted that this $T_C$ should be regarded as an order of magnitude estimate, as many simplifications are involved, e.g., the simulation is based on a classical spin model, which neglects the quantum nature of spins; the system is metallic, so the low-energy sector actually involves not only spins, but also electronic excitations; the exchange couplings are truncated at second neighbor and estimated with the crude energy mapping method, and so on. Nevertheless, the important message here is that the $T_C$ for monolayer CrTe$_{2}$ is expected to be quite high among the existing 2D magnetic materials. This is consistent with the recent experiment on few-layer (down to five layers) 1$T$-CrTe$_{2}$~\cite{sun2019room}, which revealed a high Curie temperature similar to the bulk, reaching about 320 K. The experimental observation indicates that the magnetism in CrTe$_{2}$ shows a quasi-2D character, namely, the intralayer coupling is much larger than the interlayer coupling. This implies that the room-temperature magnetism is likely to persist even down to the monolayer limit.

\section{Electronic Band structure}

After determining the ground-state magnetic configuration, we turn to the electronic band structures. Let us first consider the bulk 1$T$-CrTe$_{2}$. The calculated band structures with and without SOC are plotted in Fig.~\ref{fig4}. One observes that the system is metallic, consistent with the previous transport measurement~\cite{freitas2015ferromagnetism}. From the projected density of states (PDOS), the low-energy states around the Fermi level have contributions from both Cr-3$d$ and Te-5$p$ orbitals. The spin polarization at the Fermi level is about $20\%$. Notably, one observes in Fig.~\ref{fig4}(b) a nodal line along the $K$-$H$ path below the Fermi level around $-0.8$ eV, which is protected by the $C_{3 z}$ symmetry, {namely, the two degenerate bands on the nodal line correspond to the 2D irreducible representation of $C_{3z}$.} The degeneracy of the nodal line is lifted when the SOC is turned on.

\begin{figure*}[htbp]
	\includegraphics[width=17.8cm]{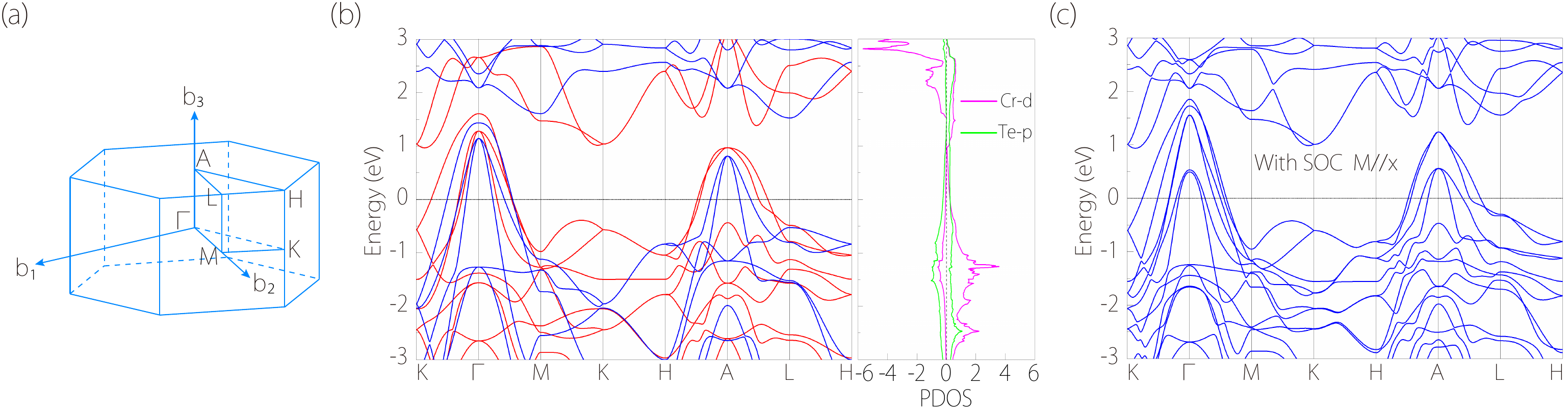}
	\caption{(a) Brillouin zone for the bulk 1$T$-CrTe$_{2}$. (b) Band structure of bulk 1$T$-CrTe$_{2}$ without SOC (left panel). The red and blue bands are for spin-up and spin-down channels, respectively. The right panel shows the spin-resolved projected density of states (PDOS). (c) Band structure of bulk 1$T$-CrTe$_{2}$ in the presence of SOC.}
	\label{fig4}
\end{figure*}

Next, we consider the monolayer CrTe$_{2}$. The corresponding band structure results are plotted in Fig.~\ref{fig5}. One observes that the system remains metallic, and the shape of the bands is also similar to that of the bulk (for paths parallel to the layer plane). Parallel to the nodal line in the bulk, one observes in Fig.~\ref{fig5}(a) a twofold degenerate Weyl point $W$ at $K$ with energy of $-0.6$ eV below the Fermi level (there is also another one at $K'$). {Here we follow the standard notions in the field of topological matters: ``Weyl" refers to twofold degeneracies, whereas ``Dirac" refers to fourfold degeneracies.} This Weyl point belongs to the majority spin channel (taken to be spin up here), hence is fully spin polarized. Note that in the absence of SOC, spin is decoupled from the spatial degree of freedom. Hence, each \emph{individual} spin channel can be \emph{effectively} considered as a spinless system, with all the original (nonmagnetic) lattice symmetries preserved, including the time reversal symmetry~\cite{wang2016time}. We find that the Weyl point $W$ is symmetry protected. {The two degenerate states at $W$ correspond to the 2D irreducible representation of the $D_3$ little group at $K$ ($K^{\prime}$). The $D_3$ group contains two generators: $C_{3z}$ and $C_{2x}$.} Subjected to these symmetries, the constraints on the effective Hamiltonian at $K$ ($K^{\prime}$) are given by
\begin{equation}
C_{3 z} \mathcal{H}_{0}\left(q_{+}, q_{-}\right) C_{3 z}^{-1}=\mathcal{H}_{0}\left(q_{+} e^{i 2 \pi / 3}, q_{-} e^{-i 2 \pi / 3}\right),
\end{equation}
\begin{equation}
C_{2x} \mathcal{H}_{0}\left(q_{x}, q_{y}\right) C_{2x}^{-1}=\mathcal{H}_{0}\left(q_{x},-q_{y}\right),
\end{equation}
where $\bm q$ is measured from $K$ ($K^{\prime}$) and $q_\pm= q_x \pm i q_y$.
We find the following effective Hamiltonian for the spin-polarized Weyl point:
\begin{equation}
\mathcal{H}_{0}(\boldsymbol{q})=v_{F}\left(\tau q_{x} \sigma_{x}+q_{y} \sigma_{y}\right),
\end{equation}
$v_{F}$ is the Fermi velocity, $\tau=\pm$ for the $K$ or $K^{\prime}$ point, and $\sigma_{i}$ are the Pauli matrices.
When the SOC is turned on, there is a small gap opening at this Weyl point, as shown in Fig.~\ref{fig5}(b).

\begin{figure}[htbp]
	\includegraphics[width=8.6cm]{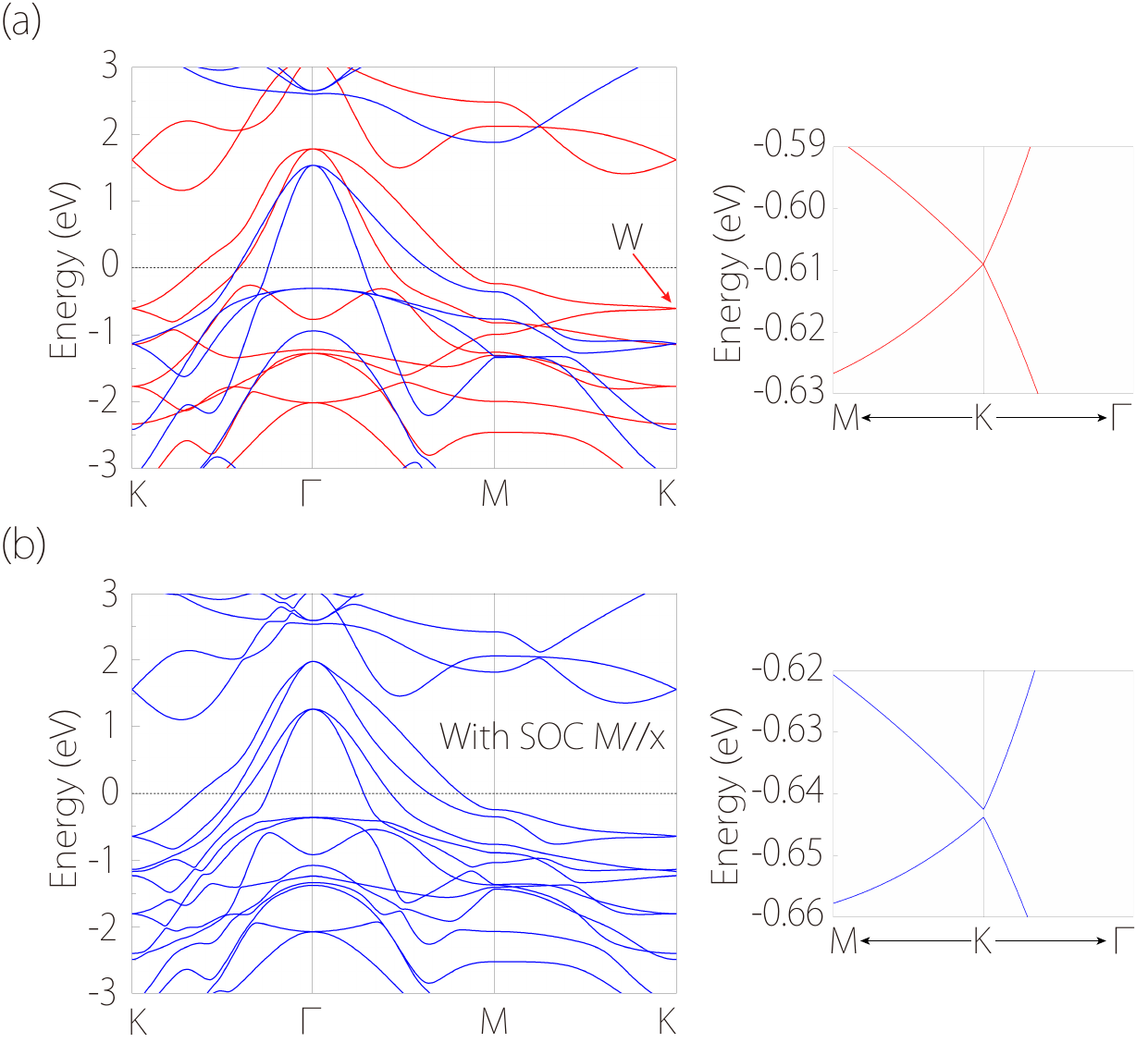}
	\caption{(a) Band structure of monolayer 1$T$-CrTe$_{2}$ without SOC (left panel). The right panel shows the zoom-in image for dispersion around the point $W$. (b) Band structure of monolayer 1$T$-CrTe$_{2}$ (with magnetization along $x$) in the presence of SOC. The right panel shows that there is a small gap opening at $W$.}
	\label{fig5}
\end{figure}

\section{Tunable Anomalous Hall effect}

The anomalous Hall effect is commonly used to characterize FM materials. The study of anomalous Hall effect has deepened our understanding of the
geometric band properties, such as the Berry phase and the Berry curvature~\cite{nagaosa2010anomalous}. Below, we will investigate this effect in bulk and monolayer
1$T$-CrTe$_{2}$.

First, we note that for both the bulk and the monolayer, the ground-state magnetic configuration preserves a vertical mirror $M_x$. This symmetry actually dictates that the in-plane anomalous Hall effect must vanish~\cite{liu2013plane}. This can be easily seen from the defining equation $j_x=\sigma_{xy}E_y$.
Under the mirror operation, $j_x$ flips sign while $E_y$ remains the same. The equation differs by an overall minus sign, thus we must have $\sigma_{xy}=0$. The argument can be readily extended to any vertical mirrors (with respect to the in-plane transport direction). Therefore, the bulk and monolayer
1$T$-CrTe$_{2}$ must have a vanishing anomalous Hall effect due to the symmetry constraint by $M_x$.

The anomalous Hall effect can be made non-vanishing by breaking the $M_x$ symmetry. In the following, we discuss two approaches to achieve this.
The most direct way is to change the magnetization direction $\hat{\bm m}$ by an applied magnetic field. The other approach is to break the symmetry via lattice strain.

To have a quantitative estimation of the anomalous Hall effect, we calculate the intrinsic anomalous Hall conductivity, which is purely determined by the band geometric properties and can be evaluated via first-principles calculations~\cite{jungwirth2002anomalous,yao2004first}. This quantity is given by
\begin{equation}
\sigma_{x y}^i=-\frac{e^{2}}{\hbar} \int_\text{BZ} \frac{d^d k}{(2\pi)^d} \Omega_z\left(\bm k\right),
\end{equation}
where $\Omega_z\left(\bm k\right)$ is the $z$-component of the total Berry curvature of the occupied states at $\bm k$,
\begin{equation}
  \Omega_{z}\left(\bm k\right)=-2 \operatorname{Im} \sum_{n\neq n^{\prime}}f_{n\bm k} \frac{\left\langle n \bm{k}\left|v_{x}\right| n' \bm{k}\right\rangle\left\langle n' \bm{k}\left|v_{y}\right| n \bm{k}\right\rangle}{(\omega_{n^{\prime}}-\omega_{n})^{2}},
\end{equation}
$n$ and $n'$ are band indices, $\varepsilon_{n}=\hbar \omega_{n}$ is the band energy, $v$'s are the velocity operators, and $f_{n\bm k}$ is the equilibrium occupation function.

Let's first consider the bulk 1$T$-CrTe$_{2}$. We have checked that for the ground state with $\hat{\bm m}$ along the $x$ direction, the intrinsic anomalous Hall conductivity vanishes identically. Then, consider the case with $\hat{\bm m}$ oriented along the $z$ direction. Figure~\ref{fig6} shows the calculated band structure and the $\sigma_{xy}^i$ as a function of the chemical potential. One observes that $\sigma_{xy}^i$ indeed becomes nonzero. $\sigma_{xy}^i$ is relatively small around the Fermi level, while pronounced peaks appear below the Fermi level.

\begin{figure}[htbp]
	\includegraphics[width=7.4cm]{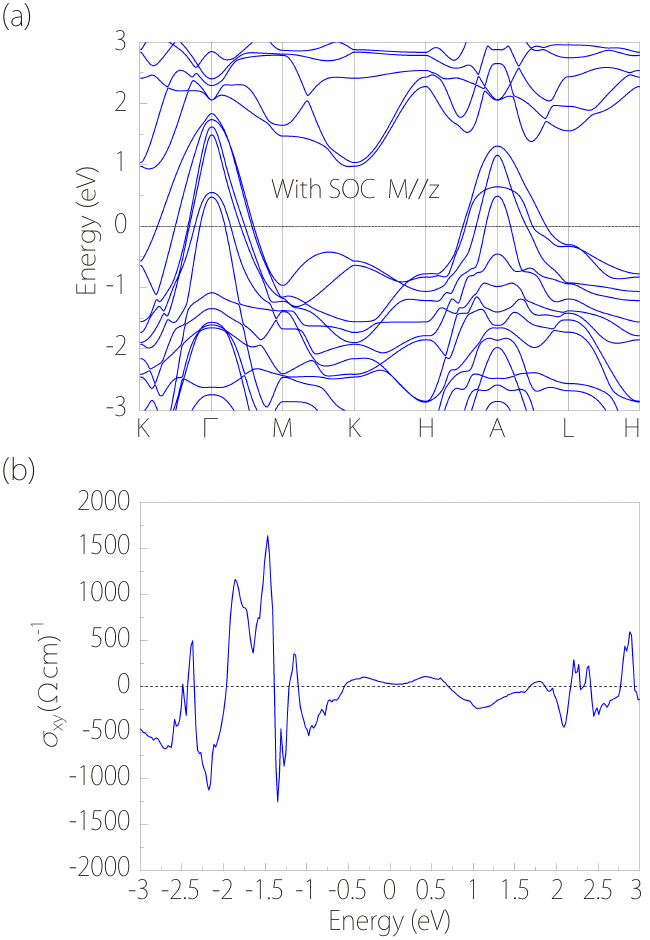}
	\caption{(a) The band structure and (b) $\sigma_{xy}^i$ versus chemical potential for monolayer 1$T$-CrTe$_{2}$ when the magnetization is along the $z$ direction. SOC is included in the calculation.}
	\label{fig6}
\end{figure}

Now, we focus on the monolayer CrTe$_{2}$, for which the magnetization can be more easily controlled. In Fig.~\ref{fig7}, we plot the band structures for $\hat{\bm m}$ along the $y$ and $z$ directions. Compared with Fig.~\ref{fig5}(b), one observes that the basic shapes of the bands are more or less the same. For magnetization along $z$, several band degeneracies are lifted at high symmetry points. Figure~\ref{fig8} shows the Berry curvature distribution for $\hat{\bm m}$ along different directions. Note that when $\hat{\bm m}\|\hat{x}$, the Berry curvature $\Omega_z$ is an odd function with respect to the mirror line perpendicular to $k_x$, indicating that its integral over the BZ vanishes due to the $M_x$ symmetry. In contrast, this symmetry is broken in Figs.~\ref{fig8}(b) and \ref{fig8}(c), so these cases have a nonzero $\sigma_{xy}^i$. As shown in Fig.~\ref{fig8}(d), $\sigma_{xy}^i$ for out-of-plane magnetization is typically larger than the in-plane case. The peak value can reach about $e^2/\hbar$, comparable to typical transition metal ferromagnets.

\begin{figure}[htbp]
	\includegraphics[width=8.8cm]{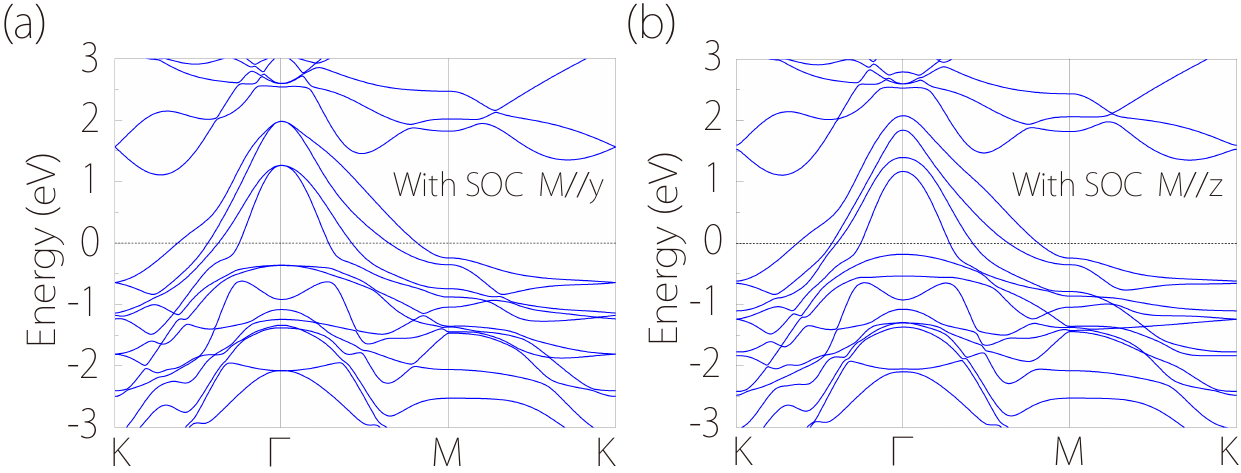}
	\caption{The band structures of monolayer 1$T$-CrTe$_{2}$ with the magnetization along (a) $y$ direction and (b) $z$ direction. SOC is included in the calculation.}
	\label{fig7}
\end{figure}

\begin{figure*}[htbp]
	\includegraphics[width=14cm]{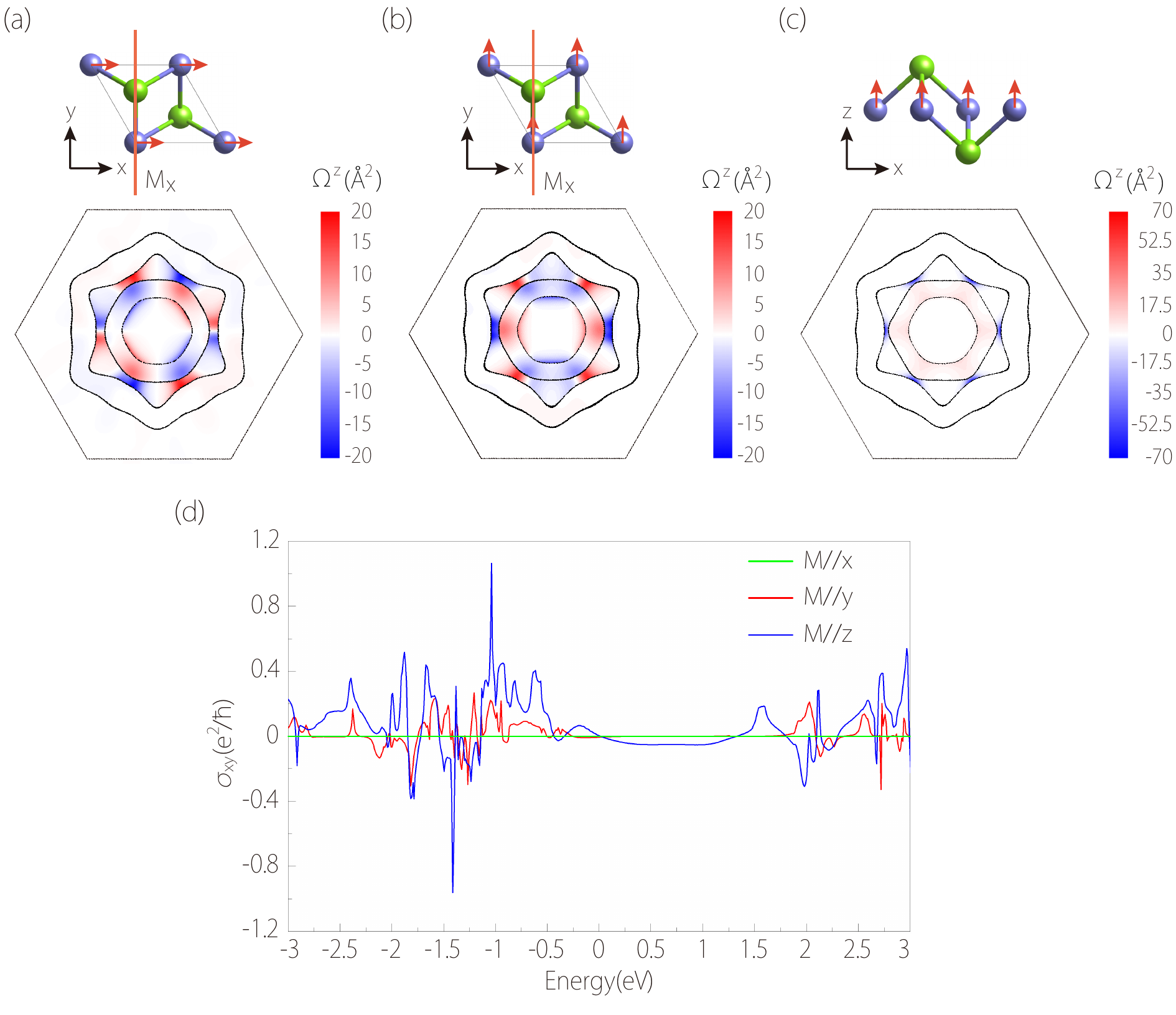}
	\caption{ Berry curvature distribution for monolayer 1$T$-CrTe$_{2}$ when the magnetization is along (a) $x$ direction, (b) $y$ direction and (c) $z$ direction (SOC included). The black lines show the Fermi contours. (d) $\sigma_{xy}^i$ plotted as a function of the chemical potential for the magnetization along different directions.}
	\label{fig8}
\end{figure*}

\begin{figure*}[htbp]
	\includegraphics[width=15cm]{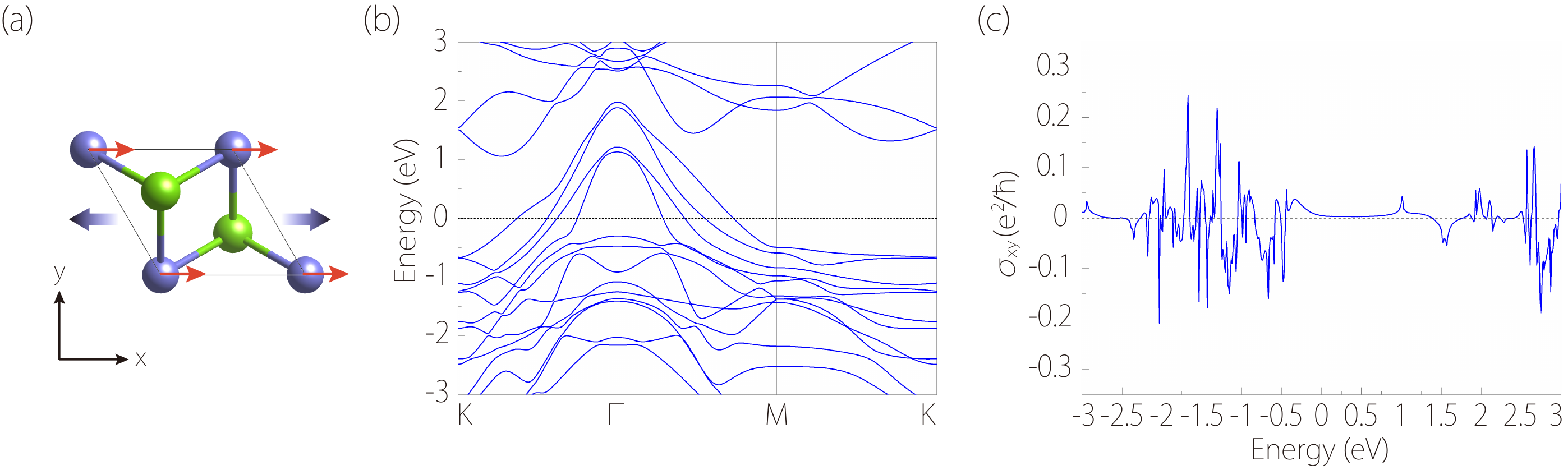}
	\caption{(a) Schematic of monolayer 1$T$-CrTe$_{2}$ under applied uniaxial strain along $x$. (b) and (c) show the calculated band structure and the intrinsic anomalous Hall conductivity at 2\% strain (SOC included).}
	\label{fig9}
\end{figure*}

Strain engineering is very powerful technique to tune the properties of 2D materials.
The vertical mirror here can also be broken by an applied uniaxial strain. Here, we consider applying the $2 \%$ uniaxial strain along the $x$ direction [see Fig.~\ref{fig9}(a)]. In this case, we find that the magnetization still prefers the $x$ direction. The energy for $\hat{\bm m}$ along the $x$ direction is lower than that along the $y$ direction by about 82 $\mu$eV per Cr, which indicates a much enhanced in-plane anisotropy. The corresponding band structure is shown in Fig.~\ref{fig9}(b). Compared to Fig.~\ref{fig5}(b), one notices that several degeneracies at the $\Gamma$ point have been lifted due to the reduction of symmetry. In fact, only the inversion symmetry $\mathcal{P}$ is preserved for the strained system. The intrinsic anomalous Hall conductivity is shown in Fig.~\ref{fig9}(c), which indeed becomes nonzero.

\section{Discussion and Conclusion}

We have a few remarks before closing. First, we have calculated the intrinsic anomalous Hall conductivity in this work. It is not the whole contribution to the anomalous Hall effect. There also exist so-called extrinsic contributions~\cite{nagaosa2010anomalous}, which originates from the scattering processes in the material, which are difficult to model accurately. Nevertheless, our calculation serves the purpose to demonstrate the dramatic tunability of the anomalous Hall effect in the material. Note that the vanishing of the anomalous Hall effect in the ground state is dictated by symmetry, which applies for both intrinsic and extrinsic contributions. We note that a recent work discussed tunable anomalous Hall effect in monolayer CrI$_3$~\cite{zhu2020theoretical}. However, for CrI$_3$, the ground state has out-of-plane ferromagnetism, with finite anomalous Hall conductivity. To turn off the anomalous Hall transport in CrI$_3$, one has to fine tune the applied magnetic field to orient the magnetic moments along certain specific in-plane directions. Thus, the switching of anomalous Hall transport in monolayer CrI$_3$ will be much more challenging compared to the proposal here.

{Second, in our DFT calculation, we take a $U$ value of 2 eV for Cr $d$ orbitals, which has been adopted in previous studies~\cite{sui2017voltage,lado2017origin}. We have also tested larger $U$ values up to 4 eV (see Supplemental Material~\cite{SM}). The result shows that the dynamic stability of the monolayer is unaffected by the $U$ value, the qualitative features of the band structure remain the same, and the larger $U$ value tends to further enhance the ferromagnetism.} Particularly, we find that the Cr magnetic moment is 3.19$\mu_B$ for $U=2$ eV, 3.32$\mu_B$ for $U=3$ eV, and 3.45$\mu_B$ for $U=4$ eV, which increases slightly with the $U$ value.

Third, to measure the anomalous Hall transport for the out-of-plane magnetization, one needs to apply a magnetic field to rotate $\hat{\bm m}$ to the $z$ direction. The applied field also generates an ordinary Hall effect, which needs to be subtracted from the measured signal. Typically, one can increase the $B$ field until the Hall voltage enters the linear-in-$B$ regime, which indicates a saturated magnetization. Then the anomalous Hall contribution can be obtained by extrapolating the linear line to the zero field limit.

{Fourth, experimentally, the Fermi level in 2D materials may be tuned by gating or chemical doping. The current experimental technique with ion liquid gating has already achieve a large doping to the order of $10^{14}$/cm$^2$ in graphene. Meanwhile, we note from Fig.~\ref{fig8}(d) that very close to the Fermi level, the anomalous Hall conductivity for out-of-plane magnetization can already reach the magnitude of $0.1$ $e^2/\hbar$, which is detectable in experiment and should exhibit a sharp contrast with in-plane magnetization.}

Finally, 2D materials typically have excellent flexibility and can sustain very large strains. For monolayer CrTe$_{2}$, we find that it can sustain a critical uniaxial strain about 19\%. The result in Fig.~\ref{fig9} is only for a small tensile strain of 2\%. One can expect that under even larger strains, the anomalous Hall effect can be further enhanced.

In conclusion, we have systematically studied the electronic, magnetic, and transport properties of bulk and monolayer 1$T$-CrTe$_{2}$.
We show that the monolayer 1$T$-CrTe$_{2}$ is a stable 2D material, which has a low exfoliation energy. The robust ferromagnetism can be maintained in the monolayer limit, with an estimated Curie temperature above $200$ K. Due to the strong SOC from the Cr-3$d$ orbital configuration, the magnetization is strongly confined in the layer plane. Both the bulk and the monolayer have ground-state magnetization perpendicular to a vertical mirror plane. The preserved mirror symmetry dictates a vanishing anomalous Hall effect for these materials. This also leads to a sensitive dependence of anomalous Hall response to external perturbations that break the symmetry. We show that re-orientation of the magnetization or a small uniaxial strain can both make a sizable anomalous Hall conductivity. Our results provide useful guidance for further studies on these interesting materials. The sensitivity of anomalous Hall transport to external field and strain could make monolayer 1$T$-CrTe$_{2}$ a promising platform for nanoscale sensors and functional devices.

\begin{acknowledgements}
The authors thank D. L. Deng for valuable discussions. This work is supported by the Singapore Ministry of Education AcRF Tier 2 (Grant No.~MOE2019-T2-1-001). S. S. Wang is supported by the Natural Science Foundation of Jiangsu Province (BK20200345) and the Research Funds for the Central Universities of China.
\end{acknowledgements}

\bibliography{FM_CrTe2_v0128}

\begin{thebibliography}{54}%
\makeatletter
\providecommand \@ifxundefined [1]{%
 \@ifx{#1\undefined}
}%
\providecommand \@ifnum [1]{%
 \ifnum #1\expandafter \@firstoftwo
 \else \expandafter \@secondoftwo
 \fi
}%
\providecommand \@ifx [1]{%
 \ifx #1\expandafter \@firstoftwo
 \else \expandafter \@secondoftwo
 \fi
}%
\providecommand \natexlab [1]{#1}%
\providecommand \enquote  [1]{``#1''}%
\providecommand \bibnamefont  [1]{#1}%
\providecommand \bibfnamefont [1]{#1}%
\providecommand \citenamefont [1]{#1}%
\providecommand \href@noop [0]{\@secondoftwo}%
\providecommand \href [0]{\begingroup \@sanitize@url \@href}%
\providecommand \@href[1]{\@@startlink{#1}\@@href}%
\providecommand \@@href[1]{\endgroup#1\@@endlink}%
\providecommand \@sanitize@url [0]{\catcode `\\12\catcode `\$12\catcode
  `\&12\catcode `\#12\catcode `\^12\catcode `\_12\catcode `\%12\relax}%
\providecommand \@@startlink[1]{}%
\providecommand \@@endlink[0]{}%
\providecommand \url  [0]{\begingroup\@sanitize@url \@url }%
\providecommand \@url [1]{\endgroup\@href {#1}{\urlprefix }}%
\providecommand \urlprefix  [0]{URL }%
\providecommand \Eprint [0]{\href }%
\providecommand \doibase [0]{http://dx.doi.org/}%
\providecommand \selectlanguage [0]{\@gobble}%
\providecommand \bibinfo  [0]{\@secondoftwo}%
\providecommand \bibfield  [0]{\@secondoftwo}%
\providecommand \translation [1]{[#1]}%
\providecommand \BibitemOpen [0]{}%
\providecommand \bibitemStop [0]{}%
\providecommand \bibitemNoStop [0]{.\EOS\space}%
\providecommand \EOS [0]{\spacefactor3000\relax}%
\providecommand \BibitemShut  [1]{\csname bibitem#1\endcsname}%
\let\auto@bib@innerbib\@empty
%</preamble>
\bibitem [{\citenamefont {Das}\ \emph {et~al.}(2015)\citenamefont {Das},
  \citenamefont {Robinson}, \citenamefont {Dubey}, \citenamefont {Terrones},\
  and\ \citenamefont {Terrones}}]{das2015beyond}%
  \BibitemOpen
  \bibfield  {author} {\bibinfo {author} {\bibfnamefont {S.}~\bibnamefont
  {Das}}, \bibinfo {author} {\bibfnamefont {J.~A.}\ \bibnamefont {Robinson}},
  \bibinfo {author} {\bibfnamefont {M.}~\bibnamefont {Dubey}}, \bibinfo
  {author} {\bibfnamefont {H.}~\bibnamefont {Terrones}}, \ and\ \bibinfo
  {author} {\bibfnamefont {M.}~\bibnamefont {Terrones}},\ }\href@noop {}
  {\bibfield  {journal} {\bibinfo  {journal} {Annu. Rev. Mater. Res.}\ }\textbf
  {\bibinfo {volume} {45}},\ \bibinfo {pages} {1} (\bibinfo {year}
  {2015})}\BibitemShut {NoStop}%
\bibitem [{\citenamefont {Bhimanapati}\ \emph {et~al.}(2015)\citenamefont
  {Bhimanapati} \emph {et~al.}}]{bhimanapati2015recent}%
  \BibitemOpen
  \bibfield  {author} {\bibinfo {author} {\bibfnamefont {G.~R.}\ \bibnamefont
  {Bhimanapati}} \emph {et~al.},\ }\href@noop {} {\bibfield  {journal}
  {\bibinfo  {journal} {ACS nano}\ }\textbf {\bibinfo {volume} {9}},\ \bibinfo
  {pages} {11509} (\bibinfo {year} {2015})}\BibitemShut {NoStop}%
\bibitem [{\citenamefont {Choi}\ \emph {et~al.}(2017)\citenamefont {Choi},
  \citenamefont {Choudhary}, \citenamefont {Han}, \citenamefont {Park},
  \citenamefont {Akinwande},\ and\ \citenamefont {Lee}}]{choi2017recent}%
  \BibitemOpen
  \bibfield  {author} {\bibinfo {author} {\bibfnamefont {W.}~\bibnamefont
  {Choi}}, \bibinfo {author} {\bibfnamefont {N.}~\bibnamefont {Choudhary}},
  \bibinfo {author} {\bibfnamefont {G.~H.}\ \bibnamefont {Han}}, \bibinfo
  {author} {\bibfnamefont {J.}~\bibnamefont {Park}}, \bibinfo {author}
  {\bibfnamefont {D.}~\bibnamefont {Akinwande}}, \ and\ \bibinfo {author}
  {\bibfnamefont {Y.~H.}\ \bibnamefont {Lee}},\ }\href@noop {} {\bibfield
  {journal} {\bibinfo  {journal} {Mater. Today}\ }\textbf {\bibinfo {volume}
  {20}},\ \bibinfo {pages} {116} (\bibinfo {year} {2017})}\BibitemShut
  {NoStop}%
\bibitem [{\citenamefont {Novoselov}\ \emph {et~al.}(2004)\citenamefont
  {Novoselov}, \citenamefont {Geim}, \citenamefont {Morozov}, \citenamefont
  {Jiang}, \citenamefont {Zhang}, \citenamefont {Dubonos}, \citenamefont
  {Grigorieva},\ and\ \citenamefont {Firsov}}]{novoselov2004electric}%
  \BibitemOpen
  \bibfield  {author} {\bibinfo {author} {\bibfnamefont {K.~S.}\ \bibnamefont
  {Novoselov}}, \bibinfo {author} {\bibfnamefont {A.~K.}\ \bibnamefont {Geim}},
  \bibinfo {author} {\bibfnamefont {S.~V.}\ \bibnamefont {Morozov}}, \bibinfo
  {author} {\bibfnamefont {D.}~\bibnamefont {Jiang}}, \bibinfo {author}
  {\bibfnamefont {Y.}~\bibnamefont {Zhang}}, \bibinfo {author} {\bibfnamefont
  {S.~V.}\ \bibnamefont {Dubonos}}, \bibinfo {author} {\bibfnamefont {I.~V.}\
  \bibnamefont {Grigorieva}}, \ and\ \bibinfo {author} {\bibfnamefont {A.~A.}\
  \bibnamefont {Firsov}},\ }\href@noop {} {\bibfield  {journal} {\bibinfo
  {journal} {science}\ }\textbf {\bibinfo {volume} {306}},\ \bibinfo {pages}
  {666} (\bibinfo {year} {2004})}\BibitemShut {NoStop}%
\bibitem [{\citenamefont {Novoselov}\ \emph {et~al.}(2005)\citenamefont
  {Novoselov}, \citenamefont {Jiang}, \citenamefont {Schedin}, \citenamefont
  {Booth}, \citenamefont {Khotkevich}, \citenamefont {Morozov},\ and\
  \citenamefont {Geim}}]{novoselov2005two}%
  \BibitemOpen
  \bibfield  {author} {\bibinfo {author} {\bibfnamefont {K.~S.}\ \bibnamefont
  {Novoselov}}, \bibinfo {author} {\bibfnamefont {D.}~\bibnamefont {Jiang}},
  \bibinfo {author} {\bibfnamefont {F.}~\bibnamefont {Schedin}}, \bibinfo
  {author} {\bibfnamefont {T.}~\bibnamefont {Booth}}, \bibinfo {author}
  {\bibfnamefont {V.}~\bibnamefont {Khotkevich}}, \bibinfo {author}
  {\bibfnamefont {S.}~\bibnamefont {Morozov}}, \ and\ \bibinfo {author}
  {\bibfnamefont {A.~K.}\ \bibnamefont {Geim}},\ }\href@noop {} {\bibfield
  {journal} {\bibinfo  {journal} {Proc. Natl. Acad. Sci. U.S.A}\ }\textbf
  {\bibinfo {volume} {102}},\ \bibinfo {pages} {10451} (\bibinfo {year}
  {2005})}\BibitemShut {NoStop}%
\bibitem [{\citenamefont {Mak}\ \emph {et~al.}(2010)\citenamefont {Mak},
  \citenamefont {Lee}, \citenamefont {Hone}, \citenamefont {Shan},\ and\
  \citenamefont {Heinz}}]{mak2010atomically}%
  \BibitemOpen
  \bibfield  {author} {\bibinfo {author} {\bibfnamefont {K.~F.}\ \bibnamefont
  {Mak}}, \bibinfo {author} {\bibfnamefont {C.}~\bibnamefont {Lee}}, \bibinfo
  {author} {\bibfnamefont {J.}~\bibnamefont {Hone}}, \bibinfo {author}
  {\bibfnamefont {J.}~\bibnamefont {Shan}}, \ and\ \bibinfo {author}
  {\bibfnamefont {T.~F.}\ \bibnamefont {Heinz}},\ }\href@noop {} {\bibfield
  {journal} {\bibinfo  {journal} {Phys. Rev. Lett.}\ }\textbf {\bibinfo
  {volume} {105}},\ \bibinfo {pages} {136805} (\bibinfo {year}
  {2010})}\BibitemShut {NoStop}%
\bibitem [{\citenamefont {Liu}\ \emph {et~al.}(2014)\citenamefont {Liu},
  \citenamefont {Neal}, \citenamefont {Zhu}, \citenamefont {Luo}, \citenamefont
  {Xu}, \citenamefont {Tom{\'a}nek},\ and\ \citenamefont
  {Ye}}]{liu2014phosphorene}%
  \BibitemOpen
  \bibfield  {author} {\bibinfo {author} {\bibfnamefont {H.}~\bibnamefont
  {Liu}}, \bibinfo {author} {\bibfnamefont {A.~T.}\ \bibnamefont {Neal}},
  \bibinfo {author} {\bibfnamefont {Z.}~\bibnamefont {Zhu}}, \bibinfo {author}
  {\bibfnamefont {Z.}~\bibnamefont {Luo}}, \bibinfo {author} {\bibfnamefont
  {X.}~\bibnamefont {Xu}}, \bibinfo {author} {\bibfnamefont {D.}~\bibnamefont
  {Tom{\'a}nek}}, \ and\ \bibinfo {author} {\bibfnamefont {P.~D.}\ \bibnamefont
  {Ye}},\ }\href@noop {} {\bibfield  {journal} {\bibinfo  {journal} {ACS nano}\
  }\textbf {\bibinfo {volume} {8}},\ \bibinfo {pages} {4033} (\bibinfo {year}
  {2014})}\BibitemShut {NoStop}%
\bibitem [{\citenamefont {Li}\ \emph {et~al.}(2014)\citenamefont {Li},
  \citenamefont {Yu}, \citenamefont {Ye}, \citenamefont {Ge}, \citenamefont
  {Ou}, \citenamefont {Wu}, \citenamefont {Feng}, \citenamefont {Chen},\ and\
  \citenamefont {Zhang}}]{li2014black}%
  \BibitemOpen
  \bibfield  {author} {\bibinfo {author} {\bibfnamefont {L.}~\bibnamefont
  {Li}}, \bibinfo {author} {\bibfnamefont {Y.}~\bibnamefont {Yu}}, \bibinfo
  {author} {\bibfnamefont {G.~J.}\ \bibnamefont {Ye}}, \bibinfo {author}
  {\bibfnamefont {Q.}~\bibnamefont {Ge}}, \bibinfo {author} {\bibfnamefont
  {X.}~\bibnamefont {Ou}}, \bibinfo {author} {\bibfnamefont {H.}~\bibnamefont
  {Wu}}, \bibinfo {author} {\bibfnamefont {D.}~\bibnamefont {Feng}}, \bibinfo
  {author} {\bibfnamefont {X.~H.}\ \bibnamefont {Chen}}, \ and\ \bibinfo
  {author} {\bibfnamefont {Y.}~\bibnamefont {Zhang}},\ }\href@noop {}
  {\bibfield  {journal} {\bibinfo  {journal} {Nat. Nanotech}\ }\textbf
  {\bibinfo {volume} {9}},\ \bibinfo {pages} {372} (\bibinfo {year}
  {2014})}\BibitemShut {NoStop}%
\bibitem [{\citenamefont {Huang}\ \emph {et~al.}(2017)\citenamefont {Huang},
  \citenamefont {Clark}, \citenamefont {Navarro-Moratalla}, \citenamefont
  {Klein}, \citenamefont {Cheng}, \citenamefont {Seyler}, \citenamefont
  {Zhong}, \citenamefont {Schmidgall}, \citenamefont {McGuire}, \citenamefont
  {Cobden}, \citenamefont {Yao}, \citenamefont {Xiao}, \citenamefont
  {Jarillo-Herrero},\ and\ \citenamefont {Xu}}]{huang2017layer}%
  \BibitemOpen
  \bibfield  {author} {\bibinfo {author} {\bibfnamefont {B.}~\bibnamefont
  {Huang}}, \bibinfo {author} {\bibfnamefont {G.}~\bibnamefont {Clark}},
  \bibinfo {author} {\bibfnamefont {E.}~\bibnamefont {Navarro-Moratalla}},
  \bibinfo {author} {\bibfnamefont {D.~R.}\ \bibnamefont {Klein}}, \bibinfo
  {author} {\bibfnamefont {R.}~\bibnamefont {Cheng}}, \bibinfo {author}
  {\bibfnamefont {K.~L.}\ \bibnamefont {Seyler}}, \bibinfo {author}
  {\bibfnamefont {D.}~\bibnamefont {Zhong}}, \bibinfo {author} {\bibfnamefont
  {E.}~\bibnamefont {Schmidgall}}, \bibinfo {author} {\bibfnamefont {M.~A.}\
  \bibnamefont {McGuire}}, \bibinfo {author} {\bibfnamefont {D.~H.}\
  \bibnamefont {Cobden}}, \bibinfo {author} {\bibfnamefont {W.}~\bibnamefont
  {Yao}}, \bibinfo {author} {\bibfnamefont {D.}~\bibnamefont {Xiao}}, \bibinfo
  {author} {\bibfnamefont {P.}~\bibnamefont {Jarillo-Herrero}}, \ and\ \bibinfo
  {author} {\bibfnamefont {X.}~\bibnamefont {Xu}},\ }\href@noop {} {\bibfield
  {journal} {\bibinfo  {journal} {Nature}\ }\textbf {\bibinfo {volume} {546}},\
  \bibinfo {pages} {270} (\bibinfo {year} {2017})}\BibitemShut {NoStop}%
\bibitem [{\citenamefont {Gong}\ \emph {et~al.}(2017)\citenamefont {Gong},
  \citenamefont {Li}, \citenamefont {Li}, \citenamefont {Ji}, \citenamefont
  {Stern}, \citenamefont {Xia}, \citenamefont {Cao}, \citenamefont {Bao},
  \citenamefont {Wang}, \citenamefont {Wang}, \citenamefont {Qiu},
  \citenamefont {Cava}, \citenamefont {Steven}, \citenamefont {Xia},\ and\
  \citenamefont {Zhang}}]{gong2017discovery}%
  \BibitemOpen
  \bibfield  {author} {\bibinfo {author} {\bibfnamefont {C.}~\bibnamefont
  {Gong}}, \bibinfo {author} {\bibfnamefont {L.}~\bibnamefont {Li}}, \bibinfo
  {author} {\bibfnamefont {Z.}~\bibnamefont {Li}}, \bibinfo {author}
  {\bibfnamefont {H.}~\bibnamefont {Ji}}, \bibinfo {author} {\bibfnamefont
  {A.}~\bibnamefont {Stern}}, \bibinfo {author} {\bibfnamefont
  {Y.}~\bibnamefont {Xia}}, \bibinfo {author} {\bibfnamefont {T.}~\bibnamefont
  {Cao}}, \bibinfo {author} {\bibfnamefont {W.}~\bibnamefont {Bao}}, \bibinfo
  {author} {\bibfnamefont {C.}~\bibnamefont {Wang}}, \bibinfo {author}
  {\bibfnamefont {Y.}~\bibnamefont {Wang}}, \bibinfo {author} {\bibfnamefont
  {Z.~Q.}\ \bibnamefont {Qiu}}, \bibinfo {author} {\bibfnamefont {R.~J.}\
  \bibnamefont {Cava}}, \bibinfo {author} {\bibfnamefont {G.~L.}\ \bibnamefont
  {Steven}}, \bibinfo {author} {\bibfnamefont {J.}~\bibnamefont {Xia}}, \ and\
  \bibinfo {author} {\bibfnamefont {X.}~\bibnamefont {Zhang}},\ }\href@noop {}
  {\bibfield  {journal} {\bibinfo  {journal} {Nature}\ }\textbf {\bibinfo
  {volume} {546}},\ \bibinfo {pages} {265} (\bibinfo {year}
  {2017})}\BibitemShut {NoStop}%
\bibitem [{\citenamefont {Zhu}\ \emph {et~al.}(2016)\citenamefont {Zhu},
  \citenamefont {Janoschek}, \citenamefont {Chaves}, \citenamefont {Cezar},
  \citenamefont {Durakiewicz}, \citenamefont {Ronning}, \citenamefont {Sassa},
  \citenamefont {Mansson}, \citenamefont {Scott}, \citenamefont {Wakeham},
  \citenamefont {Eric},\ and\ \citenamefont {Thompson}}]{zhu2016electronic}%
  \BibitemOpen
  \bibfield  {author} {\bibinfo {author} {\bibfnamefont {J.-X.}\ \bibnamefont
  {Zhu}}, \bibinfo {author} {\bibfnamefont {M.}~\bibnamefont {Janoschek}},
  \bibinfo {author} {\bibfnamefont {D.}~\bibnamefont {Chaves}}, \bibinfo
  {author} {\bibfnamefont {J.}~\bibnamefont {Cezar}}, \bibinfo {author}
  {\bibfnamefont {T.}~\bibnamefont {Durakiewicz}}, \bibinfo {author}
  {\bibfnamefont {F.}~\bibnamefont {Ronning}}, \bibinfo {author} {\bibfnamefont
  {Y.}~\bibnamefont {Sassa}}, \bibinfo {author} {\bibfnamefont
  {M.}~\bibnamefont {Mansson}}, \bibinfo {author} {\bibfnamefont
  {B.}~\bibnamefont {Scott}}, \bibinfo {author} {\bibfnamefont
  {N.}~\bibnamefont {Wakeham}}, \bibinfo {author} {\bibfnamefont {D.~B.}\
  \bibnamefont {Eric}}, \ and\ \bibinfo {author} {\bibfnamefont {J.~D.}\
  \bibnamefont {Thompson}},\ }\href@noop {} {\bibfield  {journal} {\bibinfo
  {journal} {Phys. Rev. B}\ }\textbf {\bibinfo {volume} {93}},\ \bibinfo
  {pages} {144404} (\bibinfo {year} {2016})}\BibitemShut {NoStop}%
\bibitem [{\citenamefont {Zhuang}\ \emph {et~al.}(2016)\citenamefont {Zhuang},
  \citenamefont {Kent},\ and\ \citenamefont {Hennig}}]{zhuang2016strong}%
  \BibitemOpen
  \bibfield  {author} {\bibinfo {author} {\bibfnamefont {H.~L.}\ \bibnamefont
  {Zhuang}}, \bibinfo {author} {\bibfnamefont {P.}~\bibnamefont {Kent}}, \ and\
  \bibinfo {author} {\bibfnamefont {R.~G.}\ \bibnamefont {Hennig}},\
  }\href@noop {} {\bibfield  {journal} {\bibinfo  {journal} {Phys. Rev. B}\
  }\textbf {\bibinfo {volume} {93}},\ \bibinfo {pages} {134407} (\bibinfo
  {year} {2016})}\BibitemShut {NoStop}%
\bibitem [{\citenamefont {Deng}\ \emph {et~al.}(2018)\citenamefont {Deng},
  \citenamefont {Yu}, \citenamefont {Song}, \citenamefont {Zhang},
  \citenamefont {Wang}, \citenamefont {Sun}, \citenamefont {Yi}, \citenamefont
  {Wu}, \citenamefont {Wu}, \citenamefont {Zhu}, \citenamefont {Wang},
  \citenamefont {Chen},\ and\ \citenamefont {Zhang}}]{deng2018gate}%
  \BibitemOpen
  \bibfield  {author} {\bibinfo {author} {\bibfnamefont {Y.}~\bibnamefont
  {Deng}}, \bibinfo {author} {\bibfnamefont {Y.}~\bibnamefont {Yu}}, \bibinfo
  {author} {\bibfnamefont {Y.}~\bibnamefont {Song}}, \bibinfo {author}
  {\bibfnamefont {J.}~\bibnamefont {Zhang}}, \bibinfo {author} {\bibfnamefont
  {N.~Z.}\ \bibnamefont {Wang}}, \bibinfo {author} {\bibfnamefont
  {Z.}~\bibnamefont {Sun}}, \bibinfo {author} {\bibfnamefont {Y.}~\bibnamefont
  {Yi}}, \bibinfo {author} {\bibfnamefont {Y.~Z.}\ \bibnamefont {Wu}}, \bibinfo
  {author} {\bibfnamefont {S.}~\bibnamefont {Wu}}, \bibinfo {author}
  {\bibfnamefont {J.}~\bibnamefont {Zhu}}, \bibinfo {author} {\bibfnamefont
  {J.}~\bibnamefont {Wang}}, \bibinfo {author} {\bibfnamefont {X.~H.}\
  \bibnamefont {Chen}}, \ and\ \bibinfo {author} {\bibfnamefont
  {Y.}~\bibnamefont {Zhang}},\ }\href@noop {} {\bibfield  {journal} {\bibinfo
  {journal} {Nature}\ }\textbf {\bibinfo {volume} {563}},\ \bibinfo {pages}
  {94} (\bibinfo {year} {2018})}\BibitemShut {NoStop}%
\bibitem [{\citenamefont {Bonilla}\ \emph {et~al.}(2018)\citenamefont
  {Bonilla}, \citenamefont {Kolekar}, \citenamefont {Ma}, \citenamefont {Diaz},
  \citenamefont {Kalappattil}, \citenamefont {Das}, \citenamefont {Eggers},
  \citenamefont {Gutierrez}, \citenamefont {Phan},\ and\ \citenamefont
  {Batzill}}]{bonilla2018strong}%
  \BibitemOpen
  \bibfield  {author} {\bibinfo {author} {\bibfnamefont {M.}~\bibnamefont
  {Bonilla}}, \bibinfo {author} {\bibfnamefont {S.}~\bibnamefont {Kolekar}},
  \bibinfo {author} {\bibfnamefont {Y.}~\bibnamefont {Ma}}, \bibinfo {author}
  {\bibfnamefont {H.~C.}\ \bibnamefont {Diaz}}, \bibinfo {author}
  {\bibfnamefont {V.}~\bibnamefont {Kalappattil}}, \bibinfo {author}
  {\bibfnamefont {R.}~\bibnamefont {Das}}, \bibinfo {author} {\bibfnamefont
  {T.}~\bibnamefont {Eggers}}, \bibinfo {author} {\bibfnamefont {H.~R.}\
  \bibnamefont {Gutierrez}}, \bibinfo {author} {\bibfnamefont {M.-H.}\
  \bibnamefont {Phan}}, \ and\ \bibinfo {author} {\bibfnamefont
  {M.}~\bibnamefont {Batzill}},\ }\href@noop {} {\bibfield  {journal} {\bibinfo
   {journal} {Nat. Nanotech}\ }\textbf {\bibinfo {volume} {13}},\ \bibinfo
  {pages} {289} (\bibinfo {year} {2018})}\BibitemShut {NoStop}%
\bibitem [{\citenamefont {Sugawara}\ \emph {et~al.}(2019)\citenamefont
  {Sugawara}, \citenamefont {Nakata}, \citenamefont {Fujii}, \citenamefont
  {Nakayama}, \citenamefont {Souma}, \citenamefont {Takahashi},\ and\
  \citenamefont {Sato}}]{sugawara2019monolayer}%
  \BibitemOpen
  \bibfield  {author} {\bibinfo {author} {\bibfnamefont {K.}~\bibnamefont
  {Sugawara}}, \bibinfo {author} {\bibfnamefont {Y.}~\bibnamefont {Nakata}},
  \bibinfo {author} {\bibfnamefont {K.}~\bibnamefont {Fujii}}, \bibinfo
  {author} {\bibfnamefont {K.}~\bibnamefont {Nakayama}}, \bibinfo {author}
  {\bibfnamefont {S.}~\bibnamefont {Souma}}, \bibinfo {author} {\bibfnamefont
  {T.}~\bibnamefont {Takahashi}}, \ and\ \bibinfo {author} {\bibfnamefont
  {T.}~\bibnamefont {Sato}},\ }\href@noop {} {\bibfield  {journal} {\bibinfo
  {journal} {Phys. Rev. B}\ }\textbf {\bibinfo {volume} {99}},\ \bibinfo
  {pages} {241404} (\bibinfo {year} {2019})}\BibitemShut {NoStop}%
\bibitem [{\citenamefont {O’Hara}\ \emph {et~al.}(2018)\citenamefont
  {O’Hara}, \citenamefont {Zhu}, \citenamefont {Trout}, \citenamefont
  {Ahmed}, \citenamefont {Luo}, \citenamefont {Lee}, \citenamefont {Brenner},
  \citenamefont {Rajan}, \citenamefont {Gupta}, \citenamefont {McComb},\ and\
  \citenamefont {Kawakami}}]{o2018room}%
  \BibitemOpen
  \bibfield  {author} {\bibinfo {author} {\bibfnamefont {D.~J.}\ \bibnamefont
  {O’Hara}}, \bibinfo {author} {\bibfnamefont {T.}~\bibnamefont {Zhu}},
  \bibinfo {author} {\bibfnamefont {A.~H.}\ \bibnamefont {Trout}}, \bibinfo
  {author} {\bibfnamefont {A.~S.}\ \bibnamefont {Ahmed}}, \bibinfo {author}
  {\bibfnamefont {Y.~K.}\ \bibnamefont {Luo}}, \bibinfo {author} {\bibfnamefont
  {C.~H.}\ \bibnamefont {Lee}}, \bibinfo {author} {\bibfnamefont {M.~R.}\
  \bibnamefont {Brenner}}, \bibinfo {author} {\bibfnamefont {S.}~\bibnamefont
  {Rajan}}, \bibinfo {author} {\bibfnamefont {J.~A.}\ \bibnamefont {Gupta}},
  \bibinfo {author} {\bibfnamefont {D.~W.}\ \bibnamefont {McComb}}, \ and\
  \bibinfo {author} {\bibfnamefont {R.~K.}\ \bibnamefont {Kawakami}},\
  }\href@noop {} {\bibfield  {journal} {\bibinfo  {journal} {Nano letters}\
  }\textbf {\bibinfo {volume} {18}},\ \bibinfo {pages} {3125} (\bibinfo {year}
  {2018})}\BibitemShut {NoStop}%
\bibitem [{\citenamefont {Wang}\ \emph {et~al.}(2018)\citenamefont {Wang} \emph
  {et~al.}}]{wang2018electric}%
  \BibitemOpen
  \bibfield  {author} {\bibinfo {author} {\bibfnamefont {Z.}~\bibnamefont
  {Wang}} \emph {et~al.},\ }\href@noop {} {\bibfield  {journal} {\bibinfo
  {journal} {Nat. Nanotech}\ }\textbf {\bibinfo {volume} {13}},\ \bibinfo
  {pages} {554} (\bibinfo {year} {2018})}\BibitemShut {NoStop}%
\bibitem [{\citenamefont {Sun}\ \emph {et~al.}(2019)\citenamefont {Sun},
  \citenamefont {Yi}, \citenamefont {Song}, \citenamefont {Clark},
  \citenamefont {Huang}, \citenamefont {Shan}, \citenamefont {Wu},
  \citenamefont {Huang}, \citenamefont {Gao}, \citenamefont {Chen},
  \citenamefont {McGuire}, \citenamefont {Cao}, \citenamefont {Xiao},
  \citenamefont {Liu}, \citenamefont {Yao}, \citenamefont {Xu},\ and\
  \citenamefont {Wu}}]{sun2019giant}%
  \BibitemOpen
  \bibfield  {author} {\bibinfo {author} {\bibfnamefont {Z.}~\bibnamefont
  {Sun}}, \bibinfo {author} {\bibfnamefont {Y.}~\bibnamefont {Yi}}, \bibinfo
  {author} {\bibfnamefont {T.}~\bibnamefont {Song}}, \bibinfo {author}
  {\bibfnamefont {G.}~\bibnamefont {Clark}}, \bibinfo {author} {\bibfnamefont
  {B.}~\bibnamefont {Huang}}, \bibinfo {author} {\bibfnamefont
  {Y.}~\bibnamefont {Shan}}, \bibinfo {author} {\bibfnamefont {S.}~\bibnamefont
  {Wu}}, \bibinfo {author} {\bibfnamefont {D.}~\bibnamefont {Huang}}, \bibinfo
  {author} {\bibfnamefont {C.}~\bibnamefont {Gao}}, \bibinfo {author}
  {\bibfnamefont {Z.}~\bibnamefont {Chen}}, \bibinfo {author} {\bibfnamefont
  {M.}~\bibnamefont {McGuire}}, \bibinfo {author} {\bibfnamefont
  {T.}~\bibnamefont {Cao}}, \bibinfo {author} {\bibfnamefont {D.}~\bibnamefont
  {Xiao}}, \bibinfo {author} {\bibfnamefont {W.-T.}\ \bibnamefont {Liu}},
  \bibinfo {author} {\bibfnamefont {W.}~\bibnamefont {Yao}}, \bibinfo {author}
  {\bibfnamefont {X.}~\bibnamefont {Xu}}, \ and\ \bibinfo {author}
  {\bibfnamefont {S.}~\bibnamefont {Wu}},\ }\href@noop {} {\bibfield  {journal}
  {\bibinfo  {journal} {Nature}\ }\textbf {\bibinfo {volume} {572}},\ \bibinfo
  {pages} {497} (\bibinfo {year} {2019})}\BibitemShut {NoStop}%
\bibitem [{\citenamefont {Song}\ \emph {et~al.}(2018)\citenamefont {Song},
  \citenamefont {Cai}, \citenamefont {Tu}, \citenamefont {Zhang}, \citenamefont
  {Huang}, \citenamefont {Wilson}, \citenamefont {Seyler}, \citenamefont {Zhu},
  \citenamefont {Taniguchi}, \citenamefont {Watanabe}, \citenamefont {McGuire},
  \citenamefont {Cobden}, \citenamefont {Xiao}, \citenamefont {Yao},\ and\
  \citenamefont {Xu}}]{song2018giant}%
  \BibitemOpen
  \bibfield  {author} {\bibinfo {author} {\bibfnamefont {T.}~\bibnamefont
  {Song}}, \bibinfo {author} {\bibfnamefont {X.}~\bibnamefont {Cai}}, \bibinfo
  {author} {\bibfnamefont {M.~W.-Y.}\ \bibnamefont {Tu}}, \bibinfo {author}
  {\bibfnamefont {X.}~\bibnamefont {Zhang}}, \bibinfo {author} {\bibfnamefont
  {B.}~\bibnamefont {Huang}}, \bibinfo {author} {\bibfnamefont {N.~P.}\
  \bibnamefont {Wilson}}, \bibinfo {author} {\bibfnamefont {K.~L.}\
  \bibnamefont {Seyler}}, \bibinfo {author} {\bibfnamefont {L.}~\bibnamefont
  {Zhu}}, \bibinfo {author} {\bibfnamefont {T.}~\bibnamefont {Taniguchi}},
  \bibinfo {author} {\bibfnamefont {K.}~\bibnamefont {Watanabe}}, \bibinfo
  {author} {\bibfnamefont {M.~A.}\ \bibnamefont {McGuire}}, \bibinfo {author}
  {\bibfnamefont {D.~H.}\ \bibnamefont {Cobden}}, \bibinfo {author}
  {\bibfnamefont {D.}~\bibnamefont {Xiao}}, \bibinfo {author} {\bibfnamefont
  {W.}~\bibnamefont {Yao}}, \ and\ \bibinfo {author} {\bibfnamefont
  {X.}~\bibnamefont {Xu}},\ }\href@noop {} {\bibfield  {journal} {\bibinfo
  {journal} {Science}\ }\textbf {\bibinfo {volume} {360}},\ \bibinfo {pages}
  {1214} (\bibinfo {year} {2018})}\BibitemShut {NoStop}%
\bibitem [{\citenamefont {Wang}\ \emph {et~al.}(2019)\citenamefont {Wang} \emph
  {et~al.}}]{wang2019current}%
  \BibitemOpen
  \bibfield  {author} {\bibinfo {author} {\bibfnamefont {X.}~\bibnamefont
  {Wang}} \emph {et~al.},\ }\href@noop {} {\bibfield  {journal} {\bibinfo
  {journal} {Sci. Adv.}\ }\textbf {\bibinfo {volume} {5}},\ \bibinfo {pages}
  {eaaw8904} (\bibinfo {year} {2019})}\BibitemShut {NoStop}%
\bibitem [{\citenamefont {Freitas}\ \emph {et~al.}(2015)\citenamefont
  {Freitas}, \citenamefont {Weht}, \citenamefont {Sulpice}, \citenamefont
  {Remenyi}, \citenamefont {Strobel}, \citenamefont {Gay}, \citenamefont
  {Marcus},\ and\ \citenamefont
  {N{\'u}{\~n}ez-Regueiro}}]{freitas2015ferromagnetism}%
  \BibitemOpen
  \bibfield  {author} {\bibinfo {author} {\bibfnamefont {D.~C.}\ \bibnamefont
  {Freitas}}, \bibinfo {author} {\bibfnamefont {R.}~\bibnamefont {Weht}},
  \bibinfo {author} {\bibfnamefont {A.}~\bibnamefont {Sulpice}}, \bibinfo
  {author} {\bibfnamefont {G.}~\bibnamefont {Remenyi}}, \bibinfo {author}
  {\bibfnamefont {P.}~\bibnamefont {Strobel}}, \bibinfo {author} {\bibfnamefont
  {F.}~\bibnamefont {Gay}}, \bibinfo {author} {\bibfnamefont {J.}~\bibnamefont
  {Marcus}}, \ and\ \bibinfo {author} {\bibfnamefont {M.}~\bibnamefont
  {N{\'u}{\~n}ez-Regueiro}},\ }\href@noop {} {\bibfield  {journal} {\bibinfo
  {journal} {J. Phys.: Condens. Matter}\ }\textbf {\bibinfo {volume} {27}},\
  \bibinfo {pages} {176002} (\bibinfo {year} {2015})}\BibitemShut {NoStop}%
\bibitem [{\citenamefont {Lv}\ \emph {et~al.}(2015)\citenamefont {Lv},
  \citenamefont {Lu}, \citenamefont {Shao}, \citenamefont {Liu},\ and\
  \citenamefont {Sun}}]{lv2015strain}%
  \BibitemOpen
  \bibfield  {author} {\bibinfo {author} {\bibfnamefont {H.}~\bibnamefont
  {Lv}}, \bibinfo {author} {\bibfnamefont {W.}~\bibnamefont {Lu}}, \bibinfo
  {author} {\bibfnamefont {D.}~\bibnamefont {Shao}}, \bibinfo {author}
  {\bibfnamefont {Y.}~\bibnamefont {Liu}}, \ and\ \bibinfo {author}
  {\bibfnamefont {Y.}~\bibnamefont {Sun}},\ }\href@noop {} {\bibfield
  {journal} {\bibinfo  {journal} {Phys. Rev. B}\ }\textbf {\bibinfo {volume}
  {92}},\ \bibinfo {pages} {214419} (\bibinfo {year} {2015})}\BibitemShut
  {NoStop}%
\bibitem [{\citenamefont {Sui}\ \emph {et~al.}(2017)\citenamefont {Sui},
  \citenamefont {Hu}, \citenamefont {Wang}, \citenamefont {Gu}, \citenamefont
  {Duan},\ and\ \citenamefont {Miao}}]{sui2017voltage}%
  \BibitemOpen
  \bibfield  {author} {\bibinfo {author} {\bibfnamefont {X.}~\bibnamefont
  {Sui}}, \bibinfo {author} {\bibfnamefont {T.}~\bibnamefont {Hu}}, \bibinfo
  {author} {\bibfnamefont {J.}~\bibnamefont {Wang}}, \bibinfo {author}
  {\bibfnamefont {B.-L.}\ \bibnamefont {Gu}}, \bibinfo {author} {\bibfnamefont
  {W.}~\bibnamefont {Duan}}, \ and\ \bibinfo {author} {\bibfnamefont {M.-s.}\
  \bibnamefont {Miao}},\ }\href@noop {} {\bibfield  {journal} {\bibinfo
  {journal} {Phys. Rev. B}\ }\textbf {\bibinfo {volume} {96}},\ \bibinfo
  {pages} {041410} (\bibinfo {year} {2017})}\BibitemShut {NoStop}%
\bibitem [{\citenamefont {Sun}\ \emph {et~al.}(2020)\citenamefont {Sun},
  \citenamefont {Li}, \citenamefont {Wang}, \citenamefont {Sui}, \citenamefont
  {Zhang}, \citenamefont {Wang}, \citenamefont {Liu}, \citenamefont {Li},
  \citenamefont {Feng}, \citenamefont {Zhong} \emph {et~al.}}]{sun2019room}%
  \BibitemOpen
  \bibfield  {author} {\bibinfo {author} {\bibfnamefont {X.}~\bibnamefont
  {Sun}}, \bibinfo {author} {\bibfnamefont {W.}~\bibnamefont {Li}}, \bibinfo
  {author} {\bibfnamefont {X.}~\bibnamefont {Wang}}, \bibinfo {author}
  {\bibfnamefont {Q.}~\bibnamefont {Sui}}, \bibinfo {author} {\bibfnamefont
  {T.}~\bibnamefont {Zhang}}, \bibinfo {author} {\bibfnamefont
  {Z.}~\bibnamefont {Wang}}, \bibinfo {author} {\bibfnamefont {L.}~\bibnamefont
  {Liu}}, \bibinfo {author} {\bibfnamefont {D.}~\bibnamefont {Li}}, \bibinfo
  {author} {\bibfnamefont {S.}~\bibnamefont {Feng}}, \bibinfo {author}
  {\bibfnamefont {S.}~\bibnamefont {Zhong}},  \emph {et~al.},\ }\href@noop {}
  {\bibfield  {journal} {\bibinfo  {journal} {Nano Research}\ }\textbf
  {\bibinfo {volume} {13}},\ \bibinfo {pages} {3358} (\bibinfo {year}
  {2020})}\BibitemShut {NoStop}%
\bibitem [{\citenamefont {Kresse}\ and\ \citenamefont
  {Hafner}(1994)}]{Kresse1994}%
  \BibitemOpen
  \bibfield  {author} {\bibinfo {author} {\bibfnamefont {G.}~\bibnamefont
  {Kresse}}\ and\ \bibinfo {author} {\bibfnamefont {J.}~\bibnamefont
  {Hafner}},\ }\href@noop {} {\bibfield  {journal} {\bibinfo  {journal} {Phys.
  Rev. B}\ }\textbf {\bibinfo {volume} {49}},\ \bibinfo {pages} {14251}
  (\bibinfo {year} {1994})}\BibitemShut {NoStop}%
\bibitem [{\citenamefont {Kresse}\ and\ \citenamefont
  {Furthm\"uller}(1996)}]{Kresse1996}%
  \BibitemOpen
  \bibfield  {author} {\bibinfo {author} {\bibfnamefont {G.}~\bibnamefont
  {Kresse}}\ and\ \bibinfo {author} {\bibfnamefont {J.}~\bibnamefont
  {Furthm\"uller}},\ }\href@noop {} {\bibfield  {journal} {\bibinfo  {journal}
  {Phys. Rev. B}\ }\textbf {\bibinfo {volume} {54}},\ \bibinfo {pages} {11169}
  (\bibinfo {year} {1996})}\BibitemShut {NoStop}%
\bibitem [{\citenamefont {Bl\"ochl}(1994)}]{PAW}%
  \BibitemOpen
  \bibfield  {author} {\bibinfo {author} {\bibfnamefont {P.~E.}\ \bibnamefont
  {Bl\"ochl}},\ }\href@noop {} {\bibfield  {journal} {\bibinfo  {journal}
  {Phys. Rev. B}\ }\textbf {\bibinfo {volume} {50}},\ \bibinfo {pages} {17953}
  (\bibinfo {year} {1994})}\BibitemShut {NoStop}%
\bibitem [{\citenamefont {Perdew}\ \emph {et~al.}(1996)\citenamefont {Perdew},
  \citenamefont {Burke},\ and\ \citenamefont {Ernzerhof}}]{PBE}%
  \BibitemOpen
  \bibfield  {author} {\bibinfo {author} {\bibfnamefont {J.~P.}\ \bibnamefont
  {Perdew}}, \bibinfo {author} {\bibfnamefont {K.}~\bibnamefont {Burke}}, \
  and\ \bibinfo {author} {\bibfnamefont {M.}~\bibnamefont {Ernzerhof}},\
  }\href@noop {} {\bibfield  {journal} {\bibinfo  {journal} {Phys. Rev. Lett.}\
  }\textbf {\bibinfo {volume} {77}},\ \bibinfo {pages} {3865} (\bibinfo {year}
  {1996})}\BibitemShut {NoStop}%
\bibitem [{\citenamefont {Dion}\ \emph {et~al.}(2004)\citenamefont {Dion},
  \citenamefont {Rydberg}, \citenamefont {Schr\"oder}, \citenamefont
  {Langreth},\ and\ \citenamefont {Lundqvist}}]{Dion2004}%
  \BibitemOpen
  \bibfield  {author} {\bibinfo {author} {\bibfnamefont {M.}~\bibnamefont
  {Dion}}, \bibinfo {author} {\bibfnamefont {H.}~\bibnamefont {Rydberg}},
  \bibinfo {author} {\bibfnamefont {E.}~\bibnamefont {Schr\"oder}}, \bibinfo
  {author} {\bibfnamefont {D.~C.}\ \bibnamefont {Langreth}}, \ and\ \bibinfo
  {author} {\bibfnamefont {B.~I.}\ \bibnamefont {Lundqvist}},\ }\href@noop {}
  {\bibfield  {journal} {\bibinfo  {journal} {Phys. Rev. Lett.}\ }\textbf
  {\bibinfo {volume} {92}},\ \bibinfo {pages} {246401} (\bibinfo {year}
  {2004})}\BibitemShut {NoStop}%
\bibitem [{\citenamefont {Togo}\ and\ \citenamefont {Tanaka}(2015)}]{Togo2015}%
  \BibitemOpen
  \bibfield  {author} {\bibinfo {author} {\bibfnamefont {A.}~\bibnamefont
  {Togo}}\ and\ \bibinfo {author} {\bibfnamefont {I.}~\bibnamefont {Tanaka}},\
  }\href@noop {} {\bibfield  {journal} {\bibinfo  {journal} {Scr. Mater.}\
  }\textbf {\bibinfo {volume} {108}},\ \bibinfo {pages} {1} (\bibinfo {year}
  {2015})}\BibitemShut {NoStop}%
\bibitem [{\citenamefont {Anisimov}\ \emph {et~al.}(1991)\citenamefont
  {Anisimov}, \citenamefont {Zaanen},\ and\ \citenamefont
  {Andersen}}]{Anisimov1991}%
  \BibitemOpen
  \bibfield  {author} {\bibinfo {author} {\bibfnamefont {V.~I.}\ \bibnamefont
  {Anisimov}}, \bibinfo {author} {\bibfnamefont {J.}~\bibnamefont {Zaanen}}, \
  and\ \bibinfo {author} {\bibfnamefont {O.~K.}\ \bibnamefont {Andersen}},\
  }\href@noop {} {\bibfield  {journal} {\bibinfo  {journal} {Phys. Rev. B}\
  }\textbf {\bibinfo {volume} {44}},\ \bibinfo {pages} {943} (\bibinfo {year}
  {1991})}\BibitemShut {NoStop}%
\bibitem [{\citenamefont {Dudarev}\ \emph {et~al.}(1998)\citenamefont
  {Dudarev}, \citenamefont {Botton}, \citenamefont {Savrasov}, \citenamefont
  {Humphreys},\ and\ \citenamefont {Sutton}}]{dudarev1998}%
  \BibitemOpen
  \bibfield  {author} {\bibinfo {author} {\bibfnamefont {S.~L.}\ \bibnamefont
  {Dudarev}}, \bibinfo {author} {\bibfnamefont {G.~A.}\ \bibnamefont {Botton}},
  \bibinfo {author} {\bibfnamefont {S.~Y.}\ \bibnamefont {Savrasov}}, \bibinfo
  {author} {\bibfnamefont {C.~J.}\ \bibnamefont {Humphreys}}, \ and\ \bibinfo
  {author} {\bibfnamefont {A.~P.}\ \bibnamefont {Sutton}},\ }\href@noop {}
  {\bibfield  {journal} {\bibinfo  {journal} {Phys. Rev. B}\ }\textbf {\bibinfo
  {volume} {57}},\ \bibinfo {pages} {1505} (\bibinfo {year}
  {1998})}\BibitemShut {NoStop}%
\bibitem [{\citenamefont {Lado}\ and\ \citenamefont
  {Fern{\'a}ndez-Rossier}(2017)}]{lado2017origin}%
  \BibitemOpen
  \bibfield  {author} {\bibinfo {author} {\bibfnamefont {J.~L.}\ \bibnamefont
  {Lado}}\ and\ \bibinfo {author} {\bibfnamefont {J.}~\bibnamefont
  {Fern{\'a}ndez-Rossier}},\ }\href@noop {} {\bibfield  {journal} {\bibinfo
  {journal} {2D Mater.}\ }\textbf {\bibinfo {volume} {4}},\ \bibinfo {pages}
  {035002} (\bibinfo {year} {2017})}\BibitemShut {NoStop}%
\bibitem [{SM()}]{SM}%
  \BibitemOpen
  \href@noop {} {\bibinfo  {journal} {See Supplemental Material}\ }\BibitemShut
  {NoStop}%
\bibitem [{\citenamefont {Mostofi}\ \emph {et~al.}(2008)\citenamefont
  {Mostofi}, \citenamefont {Yates}, \citenamefont {Lee}, \citenamefont {Souza},
  \citenamefont {Vanderbilt},\ and\ \citenamefont
  {Marzari}}]{mostofi2008wannier90}%
  \BibitemOpen
\bibfield  {journal} {  }\bibfield  {author} {\bibinfo {author} {\bibfnamefont
  {A.~A.}\ \bibnamefont {Mostofi}}, \bibinfo {author} {\bibfnamefont {J.~R.}\
  \bibnamefont {Yates}}, \bibinfo {author} {\bibfnamefont {Y.-S.}\ \bibnamefont
  {Lee}}, \bibinfo {author} {\bibfnamefont {I.}~\bibnamefont {Souza}}, \bibinfo
  {author} {\bibfnamefont {D.}~\bibnamefont {Vanderbilt}}, \ and\ \bibinfo
  {author} {\bibfnamefont {N.}~\bibnamefont {Marzari}},\ }\href@noop {}
  {\bibfield  {journal} {\bibinfo  {journal} {Comput. Phys. Commun.}\ }\textbf
  {\bibinfo {volume} {178}},\ \bibinfo {pages} {685} (\bibinfo {year}
  {2008})}\BibitemShut {NoStop}%
\bibitem [{\citenamefont {Wang}\ \emph {et~al.}(2006)\citenamefont {Wang},
  \citenamefont {Yates}, \citenamefont {Souza},\ and\ \citenamefont
  {Vanderbilt}}]{wang2006ab}%
  \BibitemOpen
  \bibfield  {author} {\bibinfo {author} {\bibfnamefont {X.}~\bibnamefont
  {Wang}}, \bibinfo {author} {\bibfnamefont {J.~R.}\ \bibnamefont {Yates}},
  \bibinfo {author} {\bibfnamefont {I.}~\bibnamefont {Souza}}, \ and\ \bibinfo
  {author} {\bibfnamefont {D.}~\bibnamefont {Vanderbilt}},\ }\href@noop {}
  {\bibfield  {journal} {\bibinfo  {journal} {Phys. Rev. B}\ }\textbf {\bibinfo
  {volume} {74}},\ \bibinfo {pages} {195118} (\bibinfo {year}
  {2006})}\BibitemShut {NoStop}%
\bibitem [{\citenamefont {Zacharia}\ \emph {et~al.}(2004)\citenamefont
  {Zacharia}, \citenamefont {Ulbricht},\ and\ \citenamefont
  {Hertel}}]{zacharia2004}%
  \BibitemOpen
  \bibfield  {author} {\bibinfo {author} {\bibfnamefont {R.}~\bibnamefont
  {Zacharia}}, \bibinfo {author} {\bibfnamefont {H.}~\bibnamefont {Ulbricht}},
  \ and\ \bibinfo {author} {\bibfnamefont {T.}~\bibnamefont {Hertel}},\
  }\href@noop {} {\bibfield  {journal} {\bibinfo  {journal} {Phys. Rev. B}\
  }\textbf {\bibinfo {volume} {69}},\ \bibinfo {pages} {155406} (\bibinfo
  {year} {2004})}\BibitemShut {NoStop}%
\bibitem [{\citenamefont {Zhao}\ \emph {et~al.}(2014)\citenamefont {Zhao},
  \citenamefont {Li},\ and\ \citenamefont {Yang}}]{zhao2014}%
  \BibitemOpen
  \bibfield  {author} {\bibinfo {author} {\bibfnamefont {S.}~\bibnamefont
  {Zhao}}, \bibinfo {author} {\bibfnamefont {Z.}~\bibnamefont {Li}}, \ and\
  \bibinfo {author} {\bibfnamefont {J.}~\bibnamefont {Yang}},\ }\href@noop {}
  {\bibfield  {journal} {\bibinfo  {journal} {J. Am. Chem. Soc.}\ }\textbf
  {\bibinfo {volume} {136}},\ \bibinfo {pages} {13313} (\bibinfo {year}
  {2014})}\BibitemShut {NoStop}%
\bibitem [{\citenamefont {Guan}\ \emph {et~al.}(2015)\citenamefont {Guan},
  \citenamefont {Yang}, \citenamefont {Zhu}, \citenamefont {Hu},\ and\
  \citenamefont {Yao}}]{guan2015}%
  \BibitemOpen
  \bibfield  {author} {\bibinfo {author} {\bibfnamefont {S.}~\bibnamefont
  {Guan}}, \bibinfo {author} {\bibfnamefont {S.~A.}\ \bibnamefont {Yang}},
  \bibinfo {author} {\bibfnamefont {L.}~\bibnamefont {Zhu}}, \bibinfo {author}
  {\bibfnamefont {J.}~\bibnamefont {Hu}}, \ and\ \bibinfo {author}
  {\bibfnamefont {Y.}~\bibnamefont {Yao}},\ }\href@noop {} {\bibfield
  {journal} {\bibinfo  {journal} {Sci. Rep.}\ }\textbf {\bibinfo {volume}
  {5}},\ \bibinfo {pages} {12285} (\bibinfo {year} {2015})}\BibitemShut
  {NoStop}%
\bibitem [{\citenamefont {Daalderop}\ \emph {et~al.}(1988)\citenamefont
  {Daalderop}, \citenamefont {Kelly}, \citenamefont {Schuurmans},\ and\
  \citenamefont {Jansen}}]{daalderop1988magnetic}%
  \BibitemOpen
  \bibfield  {author} {\bibinfo {author} {\bibfnamefont {G.}~\bibnamefont
  {Daalderop}}, \bibinfo {author} {\bibfnamefont {P.}~\bibnamefont {Kelly}},
  \bibinfo {author} {\bibfnamefont {M.}~\bibnamefont {Schuurmans}}, \ and\
  \bibinfo {author} {\bibfnamefont {H.}~\bibnamefont {Jansen}},\ }\href@noop {}
  {\bibfield  {journal} {\bibinfo  {journal} {J. Phys. Colloques}\ }\textbf
  {\bibinfo {volume} {49}},\ \bibinfo {pages} {C8} (\bibinfo {year}
  {1988})}\BibitemShut {NoStop}%
\bibitem [{\citenamefont {Lehnert}\ \emph {et~al.}(2010)\citenamefont
  {Lehnert}, \citenamefont {Dennler}, \citenamefont {B{\l}o{\'n}ski},
  \citenamefont {Rusponi}, \citenamefont {Etzkorn}, \citenamefont {Moulas},
  \citenamefont {Bencok}, \citenamefont {Gambardella}, \citenamefont {Brune},\
  and\ \citenamefont {Hafner}}]{lehnert2010magnetic}%
  \BibitemOpen
  \bibfield  {author} {\bibinfo {author} {\bibfnamefont {A.}~\bibnamefont
  {Lehnert}}, \bibinfo {author} {\bibfnamefont {S.}~\bibnamefont {Dennler}},
  \bibinfo {author} {\bibfnamefont {P.}~\bibnamefont {B{\l}o{\'n}ski}},
  \bibinfo {author} {\bibfnamefont {S.}~\bibnamefont {Rusponi}}, \bibinfo
  {author} {\bibfnamefont {M.}~\bibnamefont {Etzkorn}}, \bibinfo {author}
  {\bibfnamefont {G.}~\bibnamefont {Moulas}}, \bibinfo {author} {\bibfnamefont
  {P.}~\bibnamefont {Bencok}}, \bibinfo {author} {\bibfnamefont
  {P.}~\bibnamefont {Gambardella}}, \bibinfo {author} {\bibfnamefont
  {H.}~\bibnamefont {Brune}}, \ and\ \bibinfo {author} {\bibfnamefont
  {J.}~\bibnamefont {Hafner}},\ }\href@noop {} {\bibfield  {journal} {\bibinfo
  {journal} {Phys. Rev. B}\ }\textbf {\bibinfo {volume} {82}},\ \bibinfo
  {pages} {094409} (\bibinfo {year} {2010})}\BibitemShut {NoStop}%
\bibitem [{\citenamefont {Sakuma}(1994)}]{sakuma1994first}%
  \BibitemOpen
  \bibfield  {author} {\bibinfo {author} {\bibfnamefont {A.}~\bibnamefont
  {Sakuma}},\ }\href@noop {} {\bibfield  {journal} {\bibinfo  {journal} {J.
  Phys. Soc. Jpn.}\ }\textbf {\bibinfo {volume} {63}},\ \bibinfo {pages} {3053}
  (\bibinfo {year} {1994})}\BibitemShut {NoStop}%
\bibitem [{\citenamefont {Ravindran}\ \emph {et~al.}(2001)\citenamefont
  {Ravindran}, \citenamefont {Kjekshus}, \citenamefont {Fjellv{\aa}g},
  \citenamefont {James}, \citenamefont {Nordstr{\"o}m}, \citenamefont
  {Johansson},\ and\ \citenamefont {Eriksson}}]{ravindran2001large}%
  \BibitemOpen
  \bibfield  {author} {\bibinfo {author} {\bibfnamefont {P.}~\bibnamefont
  {Ravindran}}, \bibinfo {author} {\bibfnamefont {A.}~\bibnamefont {Kjekshus}},
  \bibinfo {author} {\bibfnamefont {H.}~\bibnamefont {Fjellv{\aa}g}}, \bibinfo
  {author} {\bibfnamefont {P.}~\bibnamefont {James}}, \bibinfo {author}
  {\bibfnamefont {L.}~\bibnamefont {Nordstr{\"o}m}}, \bibinfo {author}
  {\bibfnamefont {B.}~\bibnamefont {Johansson}}, \ and\ \bibinfo {author}
  {\bibfnamefont {O.}~\bibnamefont {Eriksson}},\ }\href@noop {} {\bibfield
  {journal} {\bibinfo  {journal} {Phys. Rev. B}\ }\textbf {\bibinfo {volume}
  {63}},\ \bibinfo {pages} {144409} (\bibinfo {year} {2001})}\BibitemShut
  {NoStop}%
\bibitem [{\citenamefont {Shick}\ and\ \citenamefont
  {Mryasov}(2003)}]{shick2003coulomb}%
  \BibitemOpen
  \bibfield  {author} {\bibinfo {author} {\bibfnamefont {A.~B.}\ \bibnamefont
  {Shick}}\ and\ \bibinfo {author} {\bibfnamefont {O.~N.}\ \bibnamefont
  {Mryasov}},\ }\href@noop {} {\bibfield  {journal} {\bibinfo  {journal} {Phys.
  Rev. B}\ }\textbf {\bibinfo {volume} {67}},\ \bibinfo {pages} {172407}
  (\bibinfo {year} {2003})}\BibitemShut {NoStop}%
\bibitem [{\citenamefont {Evans}\ \emph {et~al.}(2014)\citenamefont {Evans},
  \citenamefont {Fan}, \citenamefont {Chureemart}, \citenamefont {Ostler},
  \citenamefont {Ellis},\ and\ \citenamefont {Chantrell}}]{evans2014atomistic}%
  \BibitemOpen
  \bibfield  {author} {\bibinfo {author} {\bibfnamefont {R.~F.}\ \bibnamefont
  {Evans}}, \bibinfo {author} {\bibfnamefont {W.~J.}\ \bibnamefont {Fan}},
  \bibinfo {author} {\bibfnamefont {P.}~\bibnamefont {Chureemart}}, \bibinfo
  {author} {\bibfnamefont {T.~A.}\ \bibnamefont {Ostler}}, \bibinfo {author}
  {\bibfnamefont {M.~O.}\ \bibnamefont {Ellis}}, \ and\ \bibinfo {author}
  {\bibfnamefont {R.~W.}\ \bibnamefont {Chantrell}},\ }\href@noop {} {\bibfield
   {journal} {\bibinfo  {journal} {J. Phys.: Condens. Matter}\ }\textbf
  {\bibinfo {volume} {26}},\ \bibinfo {pages} {103202} (\bibinfo {year}
  {2014})}\BibitemShut {NoStop}%
\bibitem [{\citenamefont {Liechtenstein}\ \emph {et~al.}(1987)\citenamefont
  {Liechtenstein}, \citenamefont {Katsnelson}, \citenamefont {Antropov},\ and\
  \citenamefont {Gubanov}}]{liechtenstein1987local}%
  \BibitemOpen
  \bibfield  {author} {\bibinfo {author} {\bibfnamefont {A.~I.}\ \bibnamefont
  {Liechtenstein}}, \bibinfo {author} {\bibfnamefont {M.}~\bibnamefont
  {Katsnelson}}, \bibinfo {author} {\bibfnamefont {V.}~\bibnamefont
  {Antropov}}, \ and\ \bibinfo {author} {\bibfnamefont {V.}~\bibnamefont
  {Gubanov}},\ }\href@noop {} {\bibfield  {journal} {\bibinfo  {journal} {J.
  Magn. Magn. Mater.}\ }\textbf {\bibinfo {volume} {67}},\ \bibinfo {pages}
  {65} (\bibinfo {year} {1987})}\BibitemShut {NoStop}%
\bibitem [{\citenamefont {Fischer}\ \emph {et~al.}(2009)\citenamefont
  {Fischer}, \citenamefont {D{\"a}ne}, \citenamefont {Ernst}, \citenamefont
  {Bruno}, \citenamefont {L{\"u}ders}, \citenamefont {Szotek}, \citenamefont
  {Temmerman},\ and\ \citenamefont {Hergert}}]{fischer2009exchange}%
  \BibitemOpen
  \bibfield  {author} {\bibinfo {author} {\bibfnamefont {G.}~\bibnamefont
  {Fischer}}, \bibinfo {author} {\bibfnamefont {M.}~\bibnamefont {D{\"a}ne}},
  \bibinfo {author} {\bibfnamefont {A.}~\bibnamefont {Ernst}}, \bibinfo
  {author} {\bibfnamefont {P.}~\bibnamefont {Bruno}}, \bibinfo {author}
  {\bibfnamefont {M.}~\bibnamefont {L{\"u}ders}}, \bibinfo {author}
  {\bibfnamefont {Z.}~\bibnamefont {Szotek}}, \bibinfo {author} {\bibfnamefont
  {W.}~\bibnamefont {Temmerman}}, \ and\ \bibinfo {author} {\bibfnamefont
  {W.}~\bibnamefont {Hergert}},\ }\href@noop {} {\bibfield  {journal} {\bibinfo
   {journal} {Physical Review B}\ }\textbf {\bibinfo {volume} {80}},\ \bibinfo
  {pages} {014408} (\bibinfo {year} {2009})}\BibitemShut {NoStop}%
\bibitem [{\citenamefont {Nomura}\ \emph {et~al.}(2020)\citenamefont {Nomura},
  \citenamefont {Nomoto}, \citenamefont {Hirayama},\ and\ \citenamefont
  {Arita}}]{nomura2020magnetic}%
  \BibitemOpen
  \bibfield  {author} {\bibinfo {author} {\bibfnamefont {Y.}~\bibnamefont
  {Nomura}}, \bibinfo {author} {\bibfnamefont {T.}~\bibnamefont {Nomoto}},
  \bibinfo {author} {\bibfnamefont {M.}~\bibnamefont {Hirayama}}, \ and\
  \bibinfo {author} {\bibfnamefont {R.}~\bibnamefont {Arita}},\ }\href@noop {}
  {\bibfield  {journal} {\bibinfo  {journal} {Phys. Rev. Research}\ }\textbf
  {\bibinfo {volume} {2}},\ \bibinfo {pages} {043144} (\bibinfo {year}
  {2020})}\BibitemShut {NoStop}%
\bibitem [{\citenamefont {Wang}\ \emph {et~al.}(2016)\citenamefont {Wang},
  \citenamefont {Vergniory}, \citenamefont {Kushwaha}, \citenamefont
  {Hirschberger}, \citenamefont {Chulkov}, \citenamefont {Ernst}, \citenamefont
  {Ong}, \citenamefont {Cava},\ and\ \citenamefont {Bernevig}}]{wang2016time}%
  \BibitemOpen
  \bibfield  {author} {\bibinfo {author} {\bibfnamefont {Z.}~\bibnamefont
  {Wang}}, \bibinfo {author} {\bibfnamefont {M.}~\bibnamefont {Vergniory}},
  \bibinfo {author} {\bibfnamefont {S.}~\bibnamefont {Kushwaha}}, \bibinfo
  {author} {\bibfnamefont {M.}~\bibnamefont {Hirschberger}}, \bibinfo {author}
  {\bibfnamefont {E.}~\bibnamefont {Chulkov}}, \bibinfo {author} {\bibfnamefont
  {A.}~\bibnamefont {Ernst}}, \bibinfo {author} {\bibfnamefont {N.~P.}\
  \bibnamefont {Ong}}, \bibinfo {author} {\bibfnamefont {R.~J.}\ \bibnamefont
  {Cava}}, \ and\ \bibinfo {author} {\bibfnamefont {B.~A.}\ \bibnamefont
  {Bernevig}},\ }\href@noop {} {\bibfield  {journal} {\bibinfo  {journal}
  {Phys. Rev. Lett.}\ }\textbf {\bibinfo {volume} {117}},\ \bibinfo {pages}
  {236401} (\bibinfo {year} {2016})}\BibitemShut {NoStop}%
\bibitem [{\citenamefont {Nagaosa}\ \emph {et~al.}(2010)\citenamefont
  {Nagaosa}, \citenamefont {Sinova}, \citenamefont {Onoda}, \citenamefont
  {MacDonald},\ and\ \citenamefont {Ong}}]{nagaosa2010anomalous}%
  \BibitemOpen
  \bibfield  {author} {\bibinfo {author} {\bibfnamefont {N.}~\bibnamefont
  {Nagaosa}}, \bibinfo {author} {\bibfnamefont {J.}~\bibnamefont {Sinova}},
  \bibinfo {author} {\bibfnamefont {S.}~\bibnamefont {Onoda}}, \bibinfo
  {author} {\bibfnamefont {A.~H.}\ \bibnamefont {MacDonald}}, \ and\ \bibinfo
  {author} {\bibfnamefont {N.~P.}\ \bibnamefont {Ong}},\ }\href@noop {}
  {\bibfield  {journal} {\bibinfo  {journal} {Rev. Mod. Phys.}\ }\textbf
  {\bibinfo {volume} {82}},\ \bibinfo {pages} {1539} (\bibinfo {year}
  {2010})}\BibitemShut {NoStop}%
\bibitem [{\citenamefont {Liu}\ \emph {et~al.}(2013)\citenamefont {Liu},
  \citenamefont {Hsu},\ and\ \citenamefont {Liu}}]{liu2013plane}%
  \BibitemOpen
  \bibfield  {author} {\bibinfo {author} {\bibfnamefont {X.}~\bibnamefont
  {Liu}}, \bibinfo {author} {\bibfnamefont {H.-C.}\ \bibnamefont {Hsu}}, \ and\
  \bibinfo {author} {\bibfnamefont {C.-X.}\ \bibnamefont {Liu}},\ }\href@noop
  {} {\bibfield  {journal} {\bibinfo  {journal} {Phys. Rev. Lett.}\ }\textbf
  {\bibinfo {volume} {111}},\ \bibinfo {pages} {086802} (\bibinfo {year}
  {2013})}\BibitemShut {NoStop}%
\bibitem [{\citenamefont {Jungwirth}\ \emph {et~al.}(2002)\citenamefont
  {Jungwirth}, \citenamefont {Niu},\ and\ \citenamefont
  {MacDonald}}]{jungwirth2002anomalous}%
  \BibitemOpen
  \bibfield  {author} {\bibinfo {author} {\bibfnamefont {T.}~\bibnamefont
  {Jungwirth}}, \bibinfo {author} {\bibfnamefont {Q.}~\bibnamefont {Niu}}, \
  and\ \bibinfo {author} {\bibfnamefont {A.~H.}\ \bibnamefont {MacDonald}},\
  }\href@noop {} {\bibfield  {journal} {\bibinfo  {journal} {Phys. Rev. Lett.}\
  }\textbf {\bibinfo {volume} {88}},\ \bibinfo {pages} {207208} (\bibinfo
  {year} {2002})}\BibitemShut {NoStop}%
\bibitem [{\citenamefont {Yao}\ \emph {et~al.}(2004)\citenamefont {Yao},
  \citenamefont {Kleinman}, \citenamefont {MacDonald}, \citenamefont {Sinova},
  \citenamefont {Jungwirth}, \citenamefont {Wang}, \citenamefont {Wang},\ and\
  \citenamefont {Niu}}]{yao2004first}%
  \BibitemOpen
  \bibfield  {author} {\bibinfo {author} {\bibfnamefont {Y.}~\bibnamefont
  {Yao}}, \bibinfo {author} {\bibfnamefont {L.}~\bibnamefont {Kleinman}},
  \bibinfo {author} {\bibfnamefont {A.~H.}\ \bibnamefont {MacDonald}}, \bibinfo
  {author} {\bibfnamefont {J.}~\bibnamefont {Sinova}}, \bibinfo {author}
  {\bibfnamefont {T.}~\bibnamefont {Jungwirth}}, \bibinfo {author}
  {\bibfnamefont {D.-s.}\ \bibnamefont {Wang}}, \bibinfo {author}
  {\bibfnamefont {E.}~\bibnamefont {Wang}}, \ and\ \bibinfo {author}
  {\bibfnamefont {Q.}~\bibnamefont {Niu}},\ }\href@noop {} {\bibfield
  {journal} {\bibinfo  {journal} {Phys. Rev. Lett.}\ }\textbf {\bibinfo
  {volume} {92}},\ \bibinfo {pages} {037204} (\bibinfo {year}
  {2004})}\BibitemShut {NoStop}%
\bibitem [{\citenamefont {Zhu}\ \emph {et~al.}(2020)\citenamefont {Zhu},
  \citenamefont {Yao}, \citenamefont {Jiang},\ and\ \citenamefont
  {Zheng}}]{zhu2020theoretical}%
  \BibitemOpen
  \bibfield  {author} {\bibinfo {author} {\bibfnamefont {M.}~\bibnamefont
  {Zhu}}, \bibinfo {author} {\bibfnamefont {H.}~\bibnamefont {Yao}}, \bibinfo
  {author} {\bibfnamefont {L.}~\bibnamefont {Jiang}}, \ and\ \bibinfo {author}
  {\bibfnamefont {Y.}~\bibnamefont {Zheng}},\ }\href@noop {} {\bibfield
  {journal} {\bibinfo  {journal} {Appl. Phys. Lett.}\ }\textbf {\bibinfo
  {volume} {116}},\ \bibinfo {pages} {022404} (\bibinfo {year}
  {2020})}\BibitemShut {NoStop}%
\end{thebibliography}%

\end{document}